\documentstyle[12pt,epsf]{article}

\date{May 30, 1996}
\ifx\tenrm\undefined
\newcommand{\tenrm}{\small}
\else
\renewcommand{\tenrm}{\small}
\fi
\newcommand{\be}{\begin{equation}}
\newcommand{\ee}{\end{equation}}
\newcommand{\bea}{\begin{eqnarray}}
\newcommand{\eea}{\end{eqnarray}}

\begin{document}
\title{ Evidence for $\eta'$ -- $\pi$ splitting in \\ unquenched lattice QCD}

\author{R.\ Frezzotti, M.\ Masetti, R.\ Petronzio \\
\small Dipartimento di Fisica, Universit\`a di Roma {\em Tor Vergata} \\
\small and \\
\small INFN, Sezione di Roma II \\
\small Viale della Ricerca Scientifica, 00133 Roma, Italy \\
\medskip
}

\maketitle

\begin{abstract} 
We perform an extrapolation from negative to positive flavour numbers of
full QCD lattice estimates of the $\eta'$ mass. The extrapolations are 
carried out by keeping $\rho$ and $\pi$ masses at fixed values. 
We find an $\eta'$ -- $\pi$ splitting which shows a flavour dependence 
consistent with the Witten Veneziano formula based on the $U(1)$ anomaly. 
The quantitative splitting is consistent with the estimates made in the 
quenched approximation.

\end{abstract}

\vfill

\begin{flushright}
  {\bf ROM2F-96-28 ~~~~}\\
\end{flushright}

\newpage

\section{Introduction} 

The quenched approximation in lattice QCD appears to reproduce with remarkable
accuracy most of the experimental pattern of the hadronic spectrum. Unquenched
corrections are small when the comparison with the quenched case is
performed at fixed values of the lattice spacing and of the renormalized quark
mass. These two parameters can be fixed by keeping two independent physical
quantities like the $\rho$ and $\pi$ masses fixed.
The bermion method exploits the smoothness of dynamical flavour dependence 
studied at fixed renormalized quantities to make estimates in full QCD 
which are extrapolations of results obtained at negative flavour numbers,
where fermions are replaced by ``bermions'', i.e. bosons with a fermion 
action \cite{Noi}.
This procedure has allowed to show the presence of sizeable unquenching 
effects in light and heavy pseudoscalar decay constants \cite{Noi_f}. 
A quantity where unquenching effects are expected to be dramatic is the 
$\eta'$ mass: it is degenerate with the pion mass in the quenched 
approximation but splits from it in the unquenched case. According to the 
Witten Veneziano formula 

\be m_{\eta'}^2 = m_\pi^2 +3 \frac{n_f}{N_c} m_0^2 \label{witven} \ee
where $n_f$ is the number of dynamical fermion flavours, $N_c$ is the number 
of color and $m_0$ is related to the topological susceptibility and 
constant in the limit $\frac{n_f}{N_c}\rightarrow 0$ \cite{wit,ven}. 
For negative flavour numbers the splitting is expected 
to make the $\eta'$ lighter than the pion.

In this paper we study the $n_f$ dependence of the $\eta'$ mass with the 
bermion method. We find that eq.\ \ref{witven} is verified: the splitting 
$m_0^2$ is independent from $n_f$ within the statistical errors and 
is consistent with the indirect estimate made in the quenched approximation, 
which instead {\it assumes} the validity of eq.\ \ref{witven}. 
The correlations of two flavour singlet currents involve two classes 
of diagrams: those where two quark propagators connect the two currents 
(the ``fish'' diagrams) and those which need gluons to be connected 
(the ``eyes'' diagrams).
In the quenched approximation only the fish diagrams are relevant and they
lead to degenerate masses for singlet and non singlet currents.
The eyes can be evaluated in the approximation where there are no intermediate
quark loops also in the quenched approximation: this was done in 
\cite{nippon} by using a modified wall source method. 
The calculation of the singlet correlations with the bermions provides 
directly eyes and fishes. 
The first part of this paper is devoted to the detailed discusion of the
correlations used in the calculations while the second presents our results.

\section{The bermion currents} 

The action for QCD with negative flavour number $n_f$ can be written in terms 
of $n_b = |n_f|/2$ commuting spinors $\phi_b(x)$ called 
bermions: 
\be S_b^{(n_b)}[U,\phi,\phi^\dagger] = S_G[U]+\sum_{x,y,z}\sum_{i=1}^{n_b} 
\phi_i^\dagger(x)Q(x,z)Q(z,y)\phi_i(y) 
\label{not4} \ee 
where $S_G[U]$ is the gauge action, spin and color indices are implicit 
and $Q$ is the Wilson discretization of the Dirac operator
\bea
[Q\phi_i](x) = \frac{1}{2\kappa}\gamma_5 \phi_i(x)
- \frac{1}{2}\gamma_5\sum_{\mu=0}^3
U_\mu(x)(1-\gamma_\mu)\phi_i(x+\mu) \nonumber \\
- \frac{1}{2}\gamma_5\sum_{\mu=0}^3 
U^\dagger_\mu(x-\mu)(1+\gamma_\mu)\phi_i(x-\mu) 
\eea

The presence of bermions in the functional integral and in the Monte Carlo
simulation allows to measure their correlations directly as averages over the
corresponding fields. In addition, different flavours allow to disentagle 
fish diagrams from eyes and to recombine them in the mixture appropriate for
singlet currents correlations.

We have considered the following bermion currents

\be Q_5^{ij}(x) = \phi^\dagger_i(x)\gamma_5 \phi_j(x) \label{q5} \ee

\be P_5^{ij}(x) = \sum_y \phi^\dagger_i(y) Q(y,x) \phi_j(x) \label{p5} \ee
where $i$ and $j$ are flavour indices. The correlation 
$\langle P_5^{ii}(x) P_5^{ii}(y) \rangle $
corresponds to the usual fermion correlation with flavour 
diagonal pseudoscalar currents and 
receives contributions from fish and eye diagrams.
Fish diagrams can be isolated by measuring 
the correlation of flavour off-diagonal currents
\be G_\pi(x,y) = \langle P_5^{ij}(x) P_5^{ji}(y) \rangle ~~~~~ i\neq j \ee 
and eye diagrams by measuring 
\be H(x,y) = \langle P_5^{ii}(x) P_5^{jj}(y) \rangle ~~~~~ i\neq j 
\label{h} \ee 
The combination which projects onto the singlet state is:
\be G_{\eta'}(x,y) = G_\pi(x,y) -n_fH(x,y) \ee

The same correlations with $P_5^{ij}$ replaced by the pointlike bilinear 
$Q_5^{ij}$ of eq.\ \ref{q5} get the fermion propagators replaced by 
bermion propagators, which are the inverse of the square of the 
Dirac operator $Q^{2}$.
Following the technique discussed in \cite{MdaggerM} these correlations can 
be shown to be dominated by the same lowest lying states as the $P_5$ ones. 
In Appendix A we show that the only pseudoscalar one particle states 
which contribute to the correlation
\be H_{Q^2}(x,y) = \langle Q_5^{ii}(x) Q_5^{jj}(y) \rangle ~~~~~ i\neq j 
\label{h_q2} \ee 
are flavour singlet mesons, and therefore this correlation is dominated at 
large times by the $\eta'$ without contaminations from pion states. 

In general it is convenient to calculate fish correlations by standard 
inversion of the Dirac operator with fixed origin, while eyes are 
obtained only by direct computation of the correlations between bermion 
bilinears summed over all possible origins. 
In order to improve the signal of bermion correlations, we have applied
different variations  of the ``one link integral'' technique to the 
bermion field, replacing, whenever possible, a field with its average 
in the surrounding bermion and gauge configuration. 
In Appendix B we discuss briefly the method and the relevant formulae.

\section{Results} 

Our simulations are performed on a $16^3 \times 32$ lattice 
at fixed renormalized $\pi$ and $\rho$ masses:
$m_\pi =0.46$ and $m_\rho =0.66$ or $m_\pi= 0.56$ and $m_\rho = 0.71$. 
We refer to these two cases from the value of the ratio 
$R_2=m_\pi^2/m_\rho^2$ which is equal to 0.5 and 0.6 respectively: 
the smaller its value the closer is the simulation to the classical limit. 

The main results of this paper are the evidence for a state lighter than the 
pion, shown in figure 1 
for $R_2 = 0.5$ and $n_f = -4$, and the determination of the parameter $m_0$ of eq.\ \ref{witven} 
at different values of $R_2$ and as a function of the flavour number. 
This allows to investigate on further flavour dependence of the $\eta'$ 
mass beyond the one already present in the large $N_c$ limit in 
the Witten Veneziano expression. 

The determination of the mass splitting 
$\Delta m = m_{\eta'}-m_\pi$ can be extracted from the quantity 

\be R(t) = -\frac{1}{n_f}\log[G_{\eta'}(t)/G_\pi(t)] = 
-\frac{1}{n_f}\log[1 - n_fH(t)/G_\pi(t)] \label{rdef} \ee
which grows linearly with time in the region $0 \ll t \ll \frac{T}{2}$, 
where $T$ is the total lattice size in the time direction:
\be R(t) \longrightarrow \frac{\Delta m}{n_f} t \ee

In the quenched case, by taking the limit $n_f \to 0$ and by assuming 
the validity of Witten Veneziano expression, the previous formulae reduce 
to those used in \cite{nippon}: 
\be R_0(t) = \frac{H(t)}{G_\pi(t)} \longrightarrow \frac{m_0^2}{2m_\pi} t \ee 

Figures 2 to 5 show $R(t)$ as a function of time for $R_2$
= 0.5 and 0.6 and different flavour numbers ($n_f = 0$ and $n_f = -4$).
The straight lines are the behaviour expected with the central value 
of the splitting that we extract from our data, fitting $R(t)$ to a 
straight line for $3 < t < t_{\max}$ with $5 \leq t_{\max} \leq 9$. 
Measurements were taken in a Monte Carlo story consisting of the following
basic update procedure, 1 heat bath and 3 overrelaxation steps for the gauge
part and 1 heat bath and 10 overrelaxation steps for the bermion part, repeated
for a number of sweeps ranging from 23000 to 141000.
The errors on the data points are obtained with a standard jackknife algorithm
after a binning into clusters of 500 - 1000 measurements. 
For the pure gauge case, in order to keep the advantage of direct field
correlations, we have introduced pseudofermion fields which are thermalized
only in a frozen gauge configuration to provide a Monte Carlo inversion of 
the propagator. 
In order to reach a sufficient accuracy, the auxiliary system is thermalized 
for 200 configurations and the correlation measured for 500 more.
It has proven useful to introduce many (typically 4) copies of 
pseudofermion fields in order to increase the number of possible 
flavour recombinations in the correlations of currents $P_5^{ij}$ 
and $Q_5^{ij}$. 
The auxiliary fields are introduced also in the one bermion case ($n_f = -2$) 
to provide additional non dynamical flavours which disentangle eyes from 
fish diagrams. The results are reported in tables 1 and 2
together with some details on the simulation parameters.
Figure 6 and 7 summarize our results and shows $m_0$ as a 
function of the flavour number at the two values of criticality 
that we have explored. The results are consistent 
with an independence from $n_f$ of $m_0$ and confirm
the validity of the expression of eq.\ \ref{witven}. 
The value of $m_0$ at $R_2=0.5$ is also consistent with another 
determination obtained in the quenched case \cite{nippon}. 
The authors of ref.\ \cite{nippon} have explored the value of 
$m_0$ at values of quark mass lower than ours. Their extrapolation 
to the chiral limit leads to a $\eta'$ mass lower than the experimental 
value by roughly 10\%. 

As a further check of the overall picture of singlet meson masses, we have 
also stuidied the singlet and non singlet vector meson correlations. 
In figure 8 the quantity 
\be R_V(t) = -\frac{1}{n_f}\log[G_{\omega}(t)/G_\rho(t)]\ee
which is the analogous for vector mesons of $R(t)$, is shown for $n_f = -4$ 
and $R_2 = 0.5$: it is compatible with zero, indicating that the two 
states are practically degenerate and confirms the unicity of the 
pseudoscalar singlet state in QCD.

A last remark concerns the validity of the bermion approach which can 
be monitored through the value of the $\eta'$ mass: at large bermion numbers 
the particle naively speaking becomes massles and may spoil the smoothness 
of the extrapolations from negative flavour numbers. In all calculations 
that we have performed, we have used the $\eta'$ mass as a monitor of the 
limits of validity of the method. 

\newpage

\section{Appendix A}
The bermion action is 
\be S_b^{(n_b)}[U,\phi^{\dagger},\phi] = S_G[U] + \sum_x \sum_{k=1}^{n_b} 
\phi^{(b)\dagger}Q^2\phi^{(b)}\label{bermioni}\ee 
where $\phi$ and $\phi^\dagger$ are commuting fields. 
For some formal derivations it is useful to introduce an equivalent action
\be S_f^{(-|n_f|)}[U,\tilde \psi,\tilde \psi^\dagger] = 
S_G[U] +\sum_{x,y}\sum_{f=1}^{|n_f|}
\tilde \psi^{(f)\dagger}(x)\gamma_5 Q(x,y)\tilde \psi^{(f)}(y)\label{not3}\ee 
It differs from the standard QCD action only because the 
$\tilde \psi ^\dagger$ and $\tilde \psi$ fields are commuting variables. 
We consider in the following a 
theory with at least two different bermion flavours ($n_f \leq -4$). 

The correlation 
\be H_{Q^2}(t_x) = \int d^3 x 
\langle Q_5^{ii}(x) Q_5^{jj}(0) \rangle ~~~~~ i\neq j \ee 
can be written, by integrating out the bermion fields, as 
\be H_{Q^2}(t_x) = Z^{-1}\int {\cal D}U\ Tr[\gamma_5 Q^{-2}_{xx}] 
Tr[\gamma_5 Q^{-2}_{00}] e^{-S_{eff}[U]} \label{hq2t} 
\ee 
where
\be Z = \int D[U]\; e^{-S_{eff}^{(n_f)}[U]}\label{zeta} \ee 
and 
\be S_{eff}^{(n_f)}[U] = S_G[U] + \frac{|n_f|}{2} Tr(\log Q^2[U] )
\label{seffdef}\ee

Using the relation between the bermion propagator $Q^{-2}_{xy}$ 
and the fermion propagator $M^{-1}_{xy}$ 
\be Q^{-2}_{xy} = \sum_z\gamma_5 M^{-1}_{xz} \gamma_5 M^{-1}_{zy} \ee
the correlation $H_{Q^2}(t_x)$ can be written in the form
\be 
Z^{-1} \int d^3 x \int d^4  y d^4 z \int {\cal D}U\ 
Tr[M^{-1}_{xy} \gamma_5 M^{-1}_{yx}] 
Tr[M^{-1}_{0z} \gamma_5 M^{-1}_{z0}] e^{-S_{eff}[U]} 
\ee 

This expression in terms of pseudofermionic fields 
$\tilde \psi_i$ appears as a four point Green function with two of the four 
variables integrated over the whole space-time:
\begin{eqnarray}
& \int d^3 x \; \int d^4 y d^4 z \langle 0 \vert \mbox{T} \left[
\pi_{12}(y)\sigma_{21}(0)\pi_{34}(z)\sigma_{43}(x) 
\right] \vert 0 \rangle  \nonumber \\
& \left \{ \begin{array}{c}
\pi_{i j} = \tilde \psi_i^\dagger \gamma_5 \tilde  \psi_j   \\
\sigma_{i j} = \tilde \psi_i^\dagger \tilde  \psi_j  
\end{array} \right. 
\label{newcorr}
\end{eqnarray}
where the $\tilde \psi_i$ fields entering the correlation are labelled 
with a flavour index. The distinction between flavours allows 
only the contractions represented in eq.\ \ref{hq2t}.
Each new flavour has the same mass, a standard propagator $M^{-1}$, but does 
not give rise to additional fermion loops and in general it must be 
considered quenched. However, if $|n_f| \geq 4$, the flavours 
in eq.\ \ref{newcorr} can be associated with the dynamical $\psi$ like  
fields, without introducing additional quenched auxiliary flavours. 

The final expression for the continuum case in a infinite volume is obtained
by summing over all different time orderings, by inserting complete sets 
of states in the matrix elements and by using translation invariance. 
For example the contribution to the integrals in eq.\ \ref{newcorr} 
from the region with $0 < t_y < t_z < t_x$ can be written as follows: 
\bea &\sum_{a,b,c} \int d^3 x \; \int d^4 y d^4 z \theta(t_y) \theta(t_z-t_y) 
\theta(t_x-t_z) \nonumber \\
& \langle 0 \vert  \sigma_{43} (x) \vert a, \vec{p}_a \rangle
\langle a, \vec{p}_a \vert \pi_{3 4} (z) \vert b, \vec{p}_b \rangle
\langle b,\vec{p}_b \vert  \pi_{1 2} (y) \vert c, \vec{p}_c \rangle
\langle c, \vec{p}_x \vert  \sigma_{21} (0) \vert 0 \rangle \nonumber \eea 
By using translation invariance and by integrating on spatial coordinates this 
expression reduces to 
\bea & \int dt_y dt_z \theta(t_y) \theta(t_z-t_y) \theta(t_x-t_z) 
\sum_{a,b,c} e^{-m_a t_x - (m_b-m_a) t_z - (m_c-m_b) t_y}  \times \nonumber \\
& \times
\langle 0 \vert  \sigma_{43} (0) \vert a \rangle
\langle a \vert     \pi_{3 4} (0) \vert b \rangle
\langle b \vert     \pi_{1 2} (0) \vert c \rangle
\langle c \vert  \sigma_{21} (0) \vert 0 \rangle
\nonumber \eea 
where the sum runs only on states with zero spatial momentum. 
By integrating on $t_y$ and $t_z$ we obtain 
\bea & \sum_{a,b,c} 
\left\{\frac{e^{-m_c t_x}- e^{-m_a t_x}}{(m_c-m_a)(m_c-m_b)} - 
\frac{e^{-m_b t_x}- e^{-m_a t_x}}{(m_c-m_b)(m_b-m_a)}\right\} 
\times \nonumber \\
& \times
\langle 0 \vert  \sigma_{43} (0) \vert a \rangle
\langle a \vert     \pi_{3 4} (0) \vert b \rangle
\langle b \vert     \pi_{1 2} (0) \vert c \rangle
\langle c \vert  \sigma_{21} (0) \vert 0 \rangle
\nonumber \eea
The final result for positive $t_x$, obtained by considering separately 
twelve different time ordered contributions, is 
\begin{eqnarray}
& \int d^3 x & \int d^4 y d^4 z  \langle 0 \vert 
\mbox{T}\left[ \pi_{12}(y)\sigma_{21}(0)\pi_{34}(z)\sigma_{43}(x) 
 \right] \vert 0 \rangle  = \nonumber \\
& & = \sum_{a,b,c}
\left\{\frac{e^{-m_c t_x}- e^{-m_a t_x}}{(m_c-m_a)(m_c-m_b)} - 
       \frac{e^{-m_b t_x}- e^{-m_a t_x}}{(m_c-m_b)(m_b-m_a)}\right\} \times
\nonumber \\
& & \hspace{3 cm} \times
\langle 0 \vert  \sigma_{43} (0) \vert a \rangle
\langle a \vert     \pi_{3 4} (0) \vert b \rangle
\langle b \vert     \pi_{1 2} (0) \vert c \rangle
\langle c \vert  \sigma_{21} (0) \vert 0 \rangle
\nonumber \\
& & + \frac{e^{-m_b t_x}- e^{-m_c t_x}}{m_a (m_c-m_b)} \; \;
\langle 0 \vert     \pi_{3 4} (0) \vert a \rangle
\langle a \vert  \sigma_{43} (0) \vert b \rangle
\langle b \vert     \pi_{1 2} (0) \vert c \rangle
\langle c \vert  \sigma_{21} (0) \vert 0 \rangle 
\nonumber \\
& & + \frac{e^{-m_a t_x}- e^{-m_b t_x}}{m_c (m_b-m_a)}  \; \;
\langle 0 \vert  \sigma_{43} (0) \vert a \rangle
\langle a \vert     \pi_{3 4} (0) \vert b \rangle
\langle b \vert  \sigma_{21} (0) \vert c \rangle 
\langle c \vert     \pi_{1 2} (0) \vert 0 \rangle
\nonumber \\
& & + \frac{e^{-m_c t_x}}{m_a m_b} \; \;
\langle 0 \vert     \pi_{3 4} (0) \vert a \rangle
\langle a \vert     \pi_{1 2} (0) \vert b \rangle
\langle b \vert  \sigma_{43} (0) \vert c \rangle
\langle c \vert  \sigma_{21} (0) \vert 0 \rangle 
\nonumber \\
& & + \frac{e^{-m_a t_x}}{m_c m_b} \; \;
\langle 0 \vert  \sigma_{43} (0) \vert a \rangle
\langle a \vert  \sigma_{21} (0) \vert b \rangle 
\langle b \vert     \pi_{3 4} (0) \vert c \rangle
\langle c \vert     \pi_{1 2} (0) \vert 0 \rangle
\nonumber \\
& & + \frac{e^{-m_b t_x}}{m_a m_c} \; \;
\langle 0 \vert     \pi_{3 4} (0) \vert a \rangle
\langle a \vert  \sigma_{43} (0) \vert b \rangle
\langle b \vert  \sigma_{21} (0) \vert c \rangle 
\langle c \vert     \pi_{1 2} (0) \vert 0 \rangle
 \nonumber \\
& & + 
\{ \pi_{3 4}  \leftrightarrow \pi_{1 2} \} 
\label{mammut}
\end{eqnarray}
The main difference with respect to the usual correlations is that the parity 
of the intermediate states is not determined by the one of the operator chosen 
for the correlation: both parity states contribute because of the insertion 
of the parity commuter operator $\bar \psi \gamma_5 \psi$. For large times 
the lowest lying state dominates. 
In order to determine which states dominate the correlation at large times 
we consider a subgroup $SU(2) \times SU(2)$ of the flavour symmetry 
$SU(|n_f|)$. One $SU(2)$ subgroup is related to isospin trasformations 
on the flavours 1 and 2 while the other $SU(2)$ is the isospin subgroup 
for flavours 3 and 4. The  states which can give non zero contributions 
to the matrix element products in eq.\ \ref{mammut}, classified with two 
isospin numbers and parity, are (1,0)$^\pm$, (0,1)$^\pm$, (0,0)$^\pm$, 
(1,1)$^\pm$, including both flavoured pions, corresponding to (1,0)$^-$ and 
(0,1)$^-$, and the (0,0)$^-$ flavour singlet pseudoscalar $\eta'$. 
However a careful examination of eq.\ \ref{mammut} shows that the 
only states whose masses can appear in the exponentially decaying factors are 
flavour non singlet scalar mesons (1,0)$^+$ and (0,1)$^+$, 
flavour singlet pseudoscalar (0,0)$^-$ and finally pseudoscalar states 
(1,1)$^-$ with at least two mesons, while flavoured pions never enter in 
the exponentials. The conclusion is that, if the $\eta'$ is lighter then 
scalar mesons, as is always the case in our simulations, the 
correlation $H_{Q^2}$ is dominated at large times by this flavour singlet 
state without contamination from nearest states: the pions.


\section{Appendix B}

We apply to the bermions an extension of the 
``one link integral'' technique \cite{parisi_pr}
to derive new bermionic bilinears, which have the same average value 
but smaller variance with the respect to the original ones: the meson
correlations expressed in terms of these improved bilinears, are found to 
be correspondingly much more stable. 

We define the following two points Green functions: 
\bea 
G_{Q^2}(x,y) = 
\langle Q_5^{ij}(x) Q_5^{ji}(y) \rangle ~~~ i\neq j \\
G_\pi(x,y) = 
\langle P_5^{ij}(x) P_5^{ji}(y) \rangle ~~~ i\neq j 
\label{def5_1}\eea
built from the bilinears defined in eqs. \ref{q5} and \ref{p5}. 
By integrating over the bermionic fields one obtains: 
\bea 
G_{Q^2}(x,y) &= Z^{-1}\int {\cal D}U\ Tr[Q^{-2}(y,x)\gamma_5 
Q^{-2}(x,y)\gamma_5] e^{-S_{eff}[U]} \\ 
G_\pi(x,y) &= Z^{-1}\int {\cal D}U\ Tr[Q^{-1}(y,x) 
Q^{-1}(x,y)] e^{-S_{eff}[U]} \eea 
The correlation $G_\pi$ is 
equal to the standard pion correlation. 
In the study of flavour singlet pseudoscalar states (the $\eta'$ meson) 
it is also useful to consider the correlations $H(x,y)$ and $H_{Q^2}(x,y)$ 
defined in eqs \ref{h} and \ref{h_q2}. $H(x,y)$ can be written, after 
integrating on the bermionic variables, as 
\be 
H(x,y) = Z^{-1}\int {\cal D}U\ Tr[Q^{-1}(x,x)] 
Tr[Q^{-1}(y,y)] e^{-S_{eff}[U]} 
\ee 
and it isolates the eye diagram. 

We now list the general substitution rules to obtain the bermionic 
equivalent of a fermionic bilinear operator. The first is 
\be \tilde \psi \rightarrow \phi {\mbox {~~~and~~~}} 
\tilde\psi^\dagger \rightarrow \phi^\dagger Q\gamma_5 \label{subs1} 
\ee 
which gives for example the operator $P_5(x)$. The 
alternative substitution rule 
\be \tilde \psi \rightarrow Q \phi {\mbox {~~~and~~~}} 
\tilde\psi^\dagger \rightarrow \phi^\dagger\gamma_5 \label{subs2} 
\ee 
gives an equivalent operator 
\be \tilde P_5^{ij}(x) = \sum_y \phi^\dagger_i(x) Q(x,y) \phi_j(y)  \ee
Given the bilinears 
\be P_5(x) = \tilde \psi_x^\dagger \gamma_5 \tilde \psi_x \ee 
\be Q_5(x) = \phi_x^\dagger \gamma_5 \phi_x \ee 
we define their improved partners\footnote{These definitions are valid for a general bilinear with a spinor matrix $\Gamma$.} where $\Gamma = \gamma_5$: 

\be P_5'(x) = a^{-2}[\tilde\psi^\dagger (\gamma_5 Q-aI)]_x 
\Gamma [(\gamma_5 Q-aI) \tilde\psi]_x 
+ {{N_c}\over a}\sum_\alpha\Gamma_{\alpha\alpha} \ee 

\be Q_5'(x) = A^{-2}(Q^2\Phi-A\Phi)_x^\dagger \Gamma (Q^2\Phi-A\Phi)_x 
+ {{N_c}\over A}\sum_\alpha\Gamma_{\alpha\alpha} \ee 

\be P_5''(x) = a^{-4} [\tilde\psi^\dagger(\gamma_5 Q-aI)^2]_x 
\Gamma [(\gamma_5 Q-aI)^2\tilde\psi]_x 
+ {{N_c}\over {a^4}}\sum_\alpha\Gamma_{\alpha\alpha} \ee 

\be Q_5''(x) = -A^{-3}(Q^2\Phi-A\Phi)_x^\dagger 
\Gamma (Q^4\Phi-2AQ^2\Phi+A^2\Phi)_x + A^{-3}Tr[Q^4_{xx}\Gamma] 
\ee
where 
\be a = \frac{1}{2\kappa} \ee 
 and 
\be A = \frac{1 + 16\kappa^2}{4\kappa^2} \ee

The improved operators can replace the original ones with some restrictions 
on the relative time distance: 
\be
\langle P_5(x)\rangle = 
\langle P_5'(x) \rangle = 
\langle P_5''(x) \rangle \label{bilin_1}  \ee
\be\langle P_5'(x) P_5'(y) \rangle = 
\langle P_5(x) P_5(y) \rangle 
\mbox{~~~for~} |t_y-t_x| \geq 2  \ee
\be\langle P_5''(x) P_5''(y) \rangle = 
\langle P_5(x) P_5(y) \rangle 
\mbox{~~~for~} |t_y-t_x| \geq 4 \label{mescor_1} 
\ee
and

\be
\langle Q_5\rangle = \langle Q_5'(x) \rangle = 
\langle Q_5''(x) \rangle \label{bilin_2} \ee
\be\langle Q_5'(x) Q_5'(y) \rangle = 
\langle Q_5(x) Q_5(y) \rangle 
\mbox{~~~for~} |t_y-t_x| \geq 2   \ee
\be\langle Q_5''(x) Q_5''(y) \rangle = 
\langle Q_5(x) Q_5(y) \rangle 
\mbox{~~~for~} |t_y-t_x| \geq 3 \label{mescor_2}
\ee

The above relations are obtained by suitable integration of $\tilde \psi$ 
or $\phi$ fields in single lattice point which leads to their replacement
with a function of the surrounding fields.

\newpage

\begin{table}
\begin{center}
\begin{tabular}{|| c | c | c | c | c | c | c ||}\hline 
\hline 
 $n_f$ & $n_{conf}$ & $\beta$ & $\kappa$ & $m_\pi$  & $m_\rho$  & $R_2$     \\ 
\hline 
 $-6$  &     23     & 6.785   & 0.156    & 0.477(8) & 0.67(1)   & 0.507(10) \\
 $-4$  &     47     & 6.463   & 0.158    & 0.464(4) & 0.658(8)  & 0.497(9)  \\
 $-2$  &     89     & 6.1     & 0.161    & 0.467(2) & 0.662(5)  & 0.498(5)  \\ 
 $~0$  &     92     & 5.7     & 0.165    & 0.457(3) & 0.659(8)  & 0.481(7)  \\
\hline 
 $-8$  &     30     & 6.99    & 0.153    & 0.563(5) & 0.712(10) & 0.625(8)  \\
 $-6$  &     28     & 6.71    & 0.1548   & 0.559(6) & 0.716(9)  & 0.610(8)  \\
 $-4$  &     45     & 6.4     & 0.157    & 0.552(3) & 0.710(6)  & 0.604(6)  \\ 
 $~0$  &     31     & 5.7     & 0.163    & 0.562(2) & 0.719(4)  & 0.611(9)  \\
\hline 
\end{tabular}
\end{center}
\caption{The parameters of the simulations, the matched values of $\pi$ 
and $\rho$ masses and the number of gauge configurations $n_{conf}$ 
on which meson correlations are calculated by Dirac operator inversion. }
\label{TAB_1}
\end{table}

\begin{table}
\begin{center}
\begin{tabular}{|| c | c | c | c | c | c ||}\hline 
\hline 
 $n_f$ &   $n_{meas}$  & $\beta$ & $\kappa$ & $\frac{m_{\eta'}-m_\pi}{n_f}$ & $m_0$     \\ 
\hline 
 $-6$  &     23000     & 6.785   & 0.156    & 0.035(7)                      & 0.163(12) \\
 $-4$  &    121000     & 6.463   & 0.158    & 0.036(5)                      & 0.168(8)  \\
 $-2$  & 85$\times$500 & 6.1     & 0.161    & 0.029(8)                      & 0.160(22) \\ 
 $~0$  & 90$\times$500 & 5.7     & 0.165    & 0.031(7)                      & 0.167(19) \\
\hline 
 $-8$  &     72000     & 6.99    & 0.153    & 0.0146(31)                    & 0.121(12) \\
 $-6$  &     28000     & 6.71    & 0.1548   & 0.0155(60)                    & 0.126(23) \\
 $-4$  &    141000     & 6.4     & 0.157    & 0.0167(38)                    & 0.131(14) \\ 
 $~0$  & 94$\times$500 & 5.7     & 0.163    & 0.0137(36)                    & 0.124(19) \\
\hline 
\end{tabular}
\end{center}
\caption{
Results for the $\eta' - \pi$ splitting extracted from the ratio 
$R(t)$. The total number of measurements of the eye correlation $H(t)$ is 
$n_{meas}$. For $n_f = 0$ and $n_f = -2$ this correlation was measured 500 
times on each fixed gauge configuration. In the quenched case we define 
$\lim_{n_f \rightarrow 0} \frac{m_{\eta'}-m_\pi}{n_f} = \frac{m_0^2}{2m_\pi}$.
}
\label{TAB_2}
\end{table}

\newpage

\begin{figure} 
\begin{center}
\setlength{\unitlength}{0.240900pt}
\ifx\plotpoint\undefined\newsavebox{\plotpoint}\fi
\sbox{\plotpoint}{\rule[-0.175pt]{0.350pt}{0.350pt}}%
\begin{picture}(1500,1619)(0,0)
\tenrm
\sbox{\plotpoint}{\rule[-0.175pt]{0.350pt}{0.350pt}}%
\put(264,158){\rule[-0.175pt]{282.335pt}{0.350pt}}
\put(264,158){\rule[-0.175pt]{4.818pt}{0.350pt}}
\put(242,158){\makebox(0,0)[r]{0}}
\put(1416,158){\rule[-0.175pt]{4.818pt}{0.350pt}}
\put(264,383){\rule[-0.175pt]{4.818pt}{0.350pt}}
\put(242,383){\makebox(0,0)[r]{0.2}}
\put(1416,383){\rule[-0.175pt]{4.818pt}{0.350pt}}
\put(264,607){\rule[-0.175pt]{4.818pt}{0.350pt}}
\put(242,607){\makebox(0,0)[r]{0.4}}
\put(1416,607){\rule[-0.175pt]{4.818pt}{0.350pt}}
\put(264,832){\rule[-0.175pt]{4.818pt}{0.350pt}}
\put(242,832){\makebox(0,0)[r]{0.6}}
\put(1416,832){\rule[-0.175pt]{4.818pt}{0.350pt}}
\put(264,1057){\rule[-0.175pt]{4.818pt}{0.350pt}}
\put(242,1057){\makebox(0,0)[r]{0.8}}
\put(1416,1057){\rule[-0.175pt]{4.818pt}{0.350pt}}
\put(264,1281){\rule[-0.175pt]{4.818pt}{0.350pt}}
\put(242,1281){\makebox(0,0)[r]{1}}
\put(1416,1281){\rule[-0.175pt]{4.818pt}{0.350pt}}
\put(264,1506){\rule[-0.175pt]{4.818pt}{0.350pt}}
\put(242,1506){\makebox(0,0)[r]{1.2}}
\put(1416,1506){\rule[-0.175pt]{4.818pt}{0.350pt}}
\put(329,158){\rule[-0.175pt]{0.350pt}{4.818pt}}
\put(329,113){\makebox(0,0){0}}
\put(329,1486){\rule[-0.175pt]{0.350pt}{4.818pt}}
\put(459,158){\rule[-0.175pt]{0.350pt}{4.818pt}}
\put(459,113){\makebox(0,0){2}}
\put(459,1486){\rule[-0.175pt]{0.350pt}{4.818pt}}
\put(590,158){\rule[-0.175pt]{0.350pt}{4.818pt}}
\put(590,113){\makebox(0,0){4}}
\put(590,1486){\rule[-0.175pt]{0.350pt}{4.818pt}}
\put(720,158){\rule[-0.175pt]{0.350pt}{4.818pt}}
\put(720,113){\makebox(0,0){6}}
\put(720,1486){\rule[-0.175pt]{0.350pt}{4.818pt}}
\put(850,158){\rule[-0.175pt]{0.350pt}{4.818pt}}
\put(850,113){\makebox(0,0){8}}
\put(850,1486){\rule[-0.175pt]{0.350pt}{4.818pt}}
\put(980,158){\rule[-0.175pt]{0.350pt}{4.818pt}}
\put(980,113){\makebox(0,0){10}}
\put(980,1486){\rule[-0.175pt]{0.350pt}{4.818pt}}
\put(1110,158){\rule[-0.175pt]{0.350pt}{4.818pt}}
\put(1110,113){\makebox(0,0){12}}
\put(1110,1486){\rule[-0.175pt]{0.350pt}{4.818pt}}
\put(1241,158){\rule[-0.175pt]{0.350pt}{4.818pt}}
\put(1241,113){\makebox(0,0){14}}
\put(1241,1486){\rule[-0.175pt]{0.350pt}{4.818pt}}
\put(1371,158){\rule[-0.175pt]{0.350pt}{4.818pt}}
\put(1371,113){\makebox(0,0){16}}
\put(1371,1486){\rule[-0.175pt]{0.350pt}{4.818pt}}
\put(264,158){\rule[-0.175pt]{282.335pt}{0.350pt}}
\put(1436,158){\rule[-0.175pt]{0.350pt}{324.733pt}}
\put(264,1506){\rule[-0.175pt]{282.335pt}{0.350pt}}
\put(-87,832){\makebox(0,0)[l]{\shortstack{{\Large $m_{eff}(t)$}}}}
\put(850,23){\makebox(0,0){{\Large $t$}}}
\put(264,158){\rule[-0.175pt]{0.350pt}{324.733pt}}
\put(459,1449){\makebox(0,0){$\times$}}
\put(524,1015){\makebox(0,0){$\times$}}
\put(590,806){\makebox(0,0){$\times$}}
\put(655,723){\makebox(0,0){$\times$}}
\put(720,699){\makebox(0,0){$\times$}}
\put(785,693){\makebox(0,0){$\times$}}
\put(850,687){\makebox(0,0){$\times$}}
\put(915,685){\makebox(0,0){$\times$}}
\put(980,684){\makebox(0,0){$\times$}}
\put(1045,680){\makebox(0,0){$\times$}}
\put(1110,677){\makebox(0,0){$\times$}}
\put(1176,678){\makebox(0,0){$\times$}}
\put(1241,677){\makebox(0,0){$\times$}}
\put(1306,671){\makebox(0,0){$\times$}}
\put(1371,668){\makebox(0,0){$\times$}}
\put(459,1437){\rule[-0.175pt]{0.350pt}{5.782pt}}
\put(449,1437){\rule[-0.175pt]{4.818pt}{0.350pt}}
\put(449,1461){\rule[-0.175pt]{4.818pt}{0.350pt}}
\put(524,1005){\rule[-0.175pt]{0.350pt}{4.818pt}}
\put(514,1005){\rule[-0.175pt]{4.818pt}{0.350pt}}
\put(514,1025){\rule[-0.175pt]{4.818pt}{0.350pt}}
\put(590,800){\rule[-0.175pt]{0.350pt}{3.132pt}}
\put(580,800){\rule[-0.175pt]{4.818pt}{0.350pt}}
\put(580,813){\rule[-0.175pt]{4.818pt}{0.350pt}}
\put(655,716){\rule[-0.175pt]{0.350pt}{3.373pt}}
\put(645,716){\rule[-0.175pt]{4.818pt}{0.350pt}}
\put(645,730){\rule[-0.175pt]{4.818pt}{0.350pt}}
\put(720,691){\rule[-0.175pt]{0.350pt}{3.613pt}}
\put(710,691){\rule[-0.175pt]{4.818pt}{0.350pt}}
\put(710,706){\rule[-0.175pt]{4.818pt}{0.350pt}}
\put(785,687){\rule[-0.175pt]{0.350pt}{2.891pt}}
\put(775,687){\rule[-0.175pt]{4.818pt}{0.350pt}}
\put(775,699){\rule[-0.175pt]{4.818pt}{0.350pt}}
\put(850,681){\rule[-0.175pt]{0.350pt}{2.650pt}}
\put(840,681){\rule[-0.175pt]{4.818pt}{0.350pt}}
\put(840,692){\rule[-0.175pt]{4.818pt}{0.350pt}}
\put(915,678){\rule[-0.175pt]{0.350pt}{3.132pt}}
\put(905,678){\rule[-0.175pt]{4.818pt}{0.350pt}}
\put(905,691){\rule[-0.175pt]{4.818pt}{0.350pt}}
\put(980,677){\rule[-0.175pt]{0.350pt}{3.132pt}}
\put(970,677){\rule[-0.175pt]{4.818pt}{0.350pt}}
\put(970,690){\rule[-0.175pt]{4.818pt}{0.350pt}}
\put(1045,674){\rule[-0.175pt]{0.350pt}{2.650pt}}
\put(1035,674){\rule[-0.175pt]{4.818pt}{0.350pt}}
\put(1035,685){\rule[-0.175pt]{4.818pt}{0.350pt}}
\put(1110,672){\rule[-0.175pt]{0.350pt}{2.409pt}}
\put(1100,672){\rule[-0.175pt]{4.818pt}{0.350pt}}
\put(1100,682){\rule[-0.175pt]{4.818pt}{0.350pt}}
\put(1176,672){\rule[-0.175pt]{0.350pt}{2.891pt}}
\put(1166,672){\rule[-0.175pt]{4.818pt}{0.350pt}}
\put(1166,684){\rule[-0.175pt]{4.818pt}{0.350pt}}
\put(1241,670){\rule[-0.175pt]{0.350pt}{3.373pt}}
\put(1231,670){\rule[-0.175pt]{4.818pt}{0.350pt}}
\put(1231,684){\rule[-0.175pt]{4.818pt}{0.350pt}}
\put(1306,665){\rule[-0.175pt]{0.350pt}{2.891pt}}
\put(1296,665){\rule[-0.175pt]{4.818pt}{0.350pt}}
\put(1296,677){\rule[-0.175pt]{4.818pt}{0.350pt}}
\put(1371,660){\rule[-0.175pt]{0.350pt}{3.613pt}}
\put(1361,660){\rule[-0.175pt]{4.818pt}{0.350pt}}
\put(1361,675){\rule[-0.175pt]{4.818pt}{0.350pt}}
\put(394,242){\makebox(0,0){$\star$}}
\put(459,337){\makebox(0,0){$\star$}}
\put(524,399){\makebox(0,0){$\star$}}
\put(590,434){\makebox(0,0){$\star$}}
\put(655,455){\makebox(0,0){$\star$}}
\put(720,468){\makebox(0,0){$\star$}}
\put(785,479){\makebox(0,0){$\star$}}
\put(850,491){\makebox(0,0){$\star$}}
\put(915,499){\makebox(0,0){$\star$}}
\put(980,506){\makebox(0,0){$\star$}}
\put(1045,518){\makebox(0,0){$\star$}}
\put(1110,521){\makebox(0,0){$\star$}}
\put(1176,509){\makebox(0,0){$\star$}}
\put(1241,490){\makebox(0,0){$\star$}}
\put(1306,479){\makebox(0,0){$\star$}}
\put(1371,467){\makebox(0,0){$\star$}}
\put(394,242){\usebox{\plotpoint}}
\put(384,242){\rule[-0.175pt]{4.818pt}{0.350pt}}
\put(384,243){\rule[-0.175pt]{4.818pt}{0.350pt}}
\put(459,335){\rule[-0.175pt]{0.350pt}{0.964pt}}
\put(449,335){\rule[-0.175pt]{4.818pt}{0.350pt}}
\put(449,339){\rule[-0.175pt]{4.818pt}{0.350pt}}
\put(524,395){\rule[-0.175pt]{0.350pt}{1.686pt}}
\put(514,395){\rule[-0.175pt]{4.818pt}{0.350pt}}
\put(514,402){\rule[-0.175pt]{4.818pt}{0.350pt}}
\put(590,429){\rule[-0.175pt]{0.350pt}{2.409pt}}
\put(580,429){\rule[-0.175pt]{4.818pt}{0.350pt}}
\put(580,439){\rule[-0.175pt]{4.818pt}{0.350pt}}
\put(655,448){\rule[-0.175pt]{0.350pt}{3.373pt}}
\put(645,448){\rule[-0.175pt]{4.818pt}{0.350pt}}
\put(645,462){\rule[-0.175pt]{4.818pt}{0.350pt}}
\put(720,458){\rule[-0.175pt]{0.350pt}{4.818pt}}
\put(710,458){\rule[-0.175pt]{4.818pt}{0.350pt}}
\put(710,478){\rule[-0.175pt]{4.818pt}{0.350pt}}
\put(785,466){\rule[-0.175pt]{0.350pt}{6.263pt}}
\put(775,466){\rule[-0.175pt]{4.818pt}{0.350pt}}
\put(775,492){\rule[-0.175pt]{4.818pt}{0.350pt}}
\put(850,474){\rule[-0.175pt]{0.350pt}{8.431pt}}
\put(840,474){\rule[-0.175pt]{4.818pt}{0.350pt}}
\put(840,509){\rule[-0.175pt]{4.818pt}{0.350pt}}
\put(915,475){\rule[-0.175pt]{0.350pt}{11.322pt}}
\put(905,475){\rule[-0.175pt]{4.818pt}{0.350pt}}
\put(905,522){\rule[-0.175pt]{4.818pt}{0.350pt}}
\put(980,473){\rule[-0.175pt]{0.350pt}{15.658pt}}
\put(970,473){\rule[-0.175pt]{4.818pt}{0.350pt}}
\put(970,538){\rule[-0.175pt]{4.818pt}{0.350pt}}
\put(1045,473){\rule[-0.175pt]{0.350pt}{21.922pt}}
\put(1035,473){\rule[-0.175pt]{4.818pt}{0.350pt}}
\put(1035,564){\rule[-0.175pt]{4.818pt}{0.350pt}}
\put(1110,460){\rule[-0.175pt]{0.350pt}{29.390pt}}
\put(1100,460){\rule[-0.175pt]{4.818pt}{0.350pt}}
\put(1100,582){\rule[-0.175pt]{4.818pt}{0.350pt}}
\put(1176,431){\rule[-0.175pt]{0.350pt}{37.580pt}}
\put(1166,431){\rule[-0.175pt]{4.818pt}{0.350pt}}
\put(1166,587){\rule[-0.175pt]{4.818pt}{0.350pt}}
\put(1241,397){\rule[-0.175pt]{0.350pt}{44.807pt}}
\put(1231,397){\rule[-0.175pt]{4.818pt}{0.350pt}}
\put(1231,583){\rule[-0.175pt]{4.818pt}{0.350pt}}
\put(1306,373){\rule[-0.175pt]{0.350pt}{51.071pt}}
\put(1296,373){\rule[-0.175pt]{4.818pt}{0.350pt}}
\put(1296,585){\rule[-0.175pt]{4.818pt}{0.350pt}}
\put(1371,351){\rule[-0.175pt]{0.350pt}{55.889pt}}
\put(1361,351){\rule[-0.175pt]{4.818pt}{0.350pt}}
\put(1361,583){\rule[-0.175pt]{4.818pt}{0.350pt}}
\end{picture}
\caption{$\times$: Pion effective mass from the correlation $G_\pi$
for $n_f = -4$ and $R_2 = 0.5$ ($\times$); 
flavour singlet pseudoscalar effective mass from the 
correlation $H_{Q^2}$ in the same simulation ($\star$).} 
\end{center}
\end{figure}
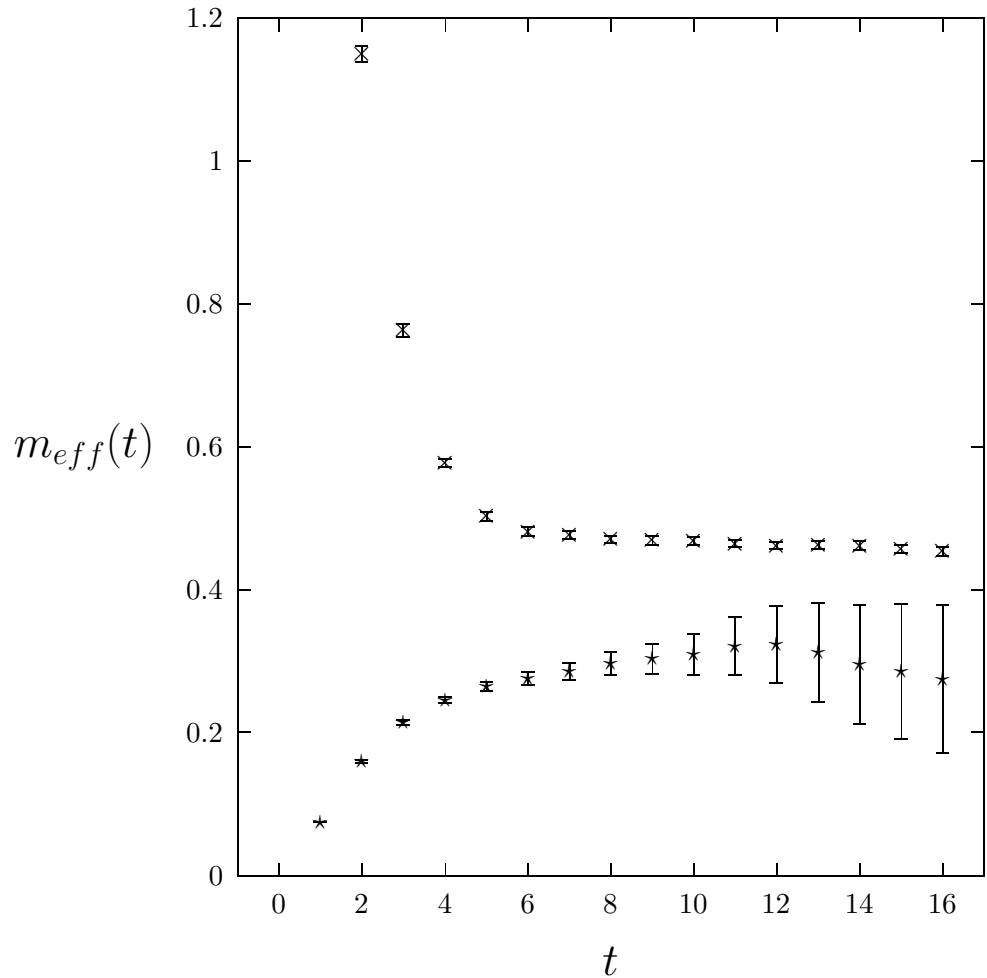

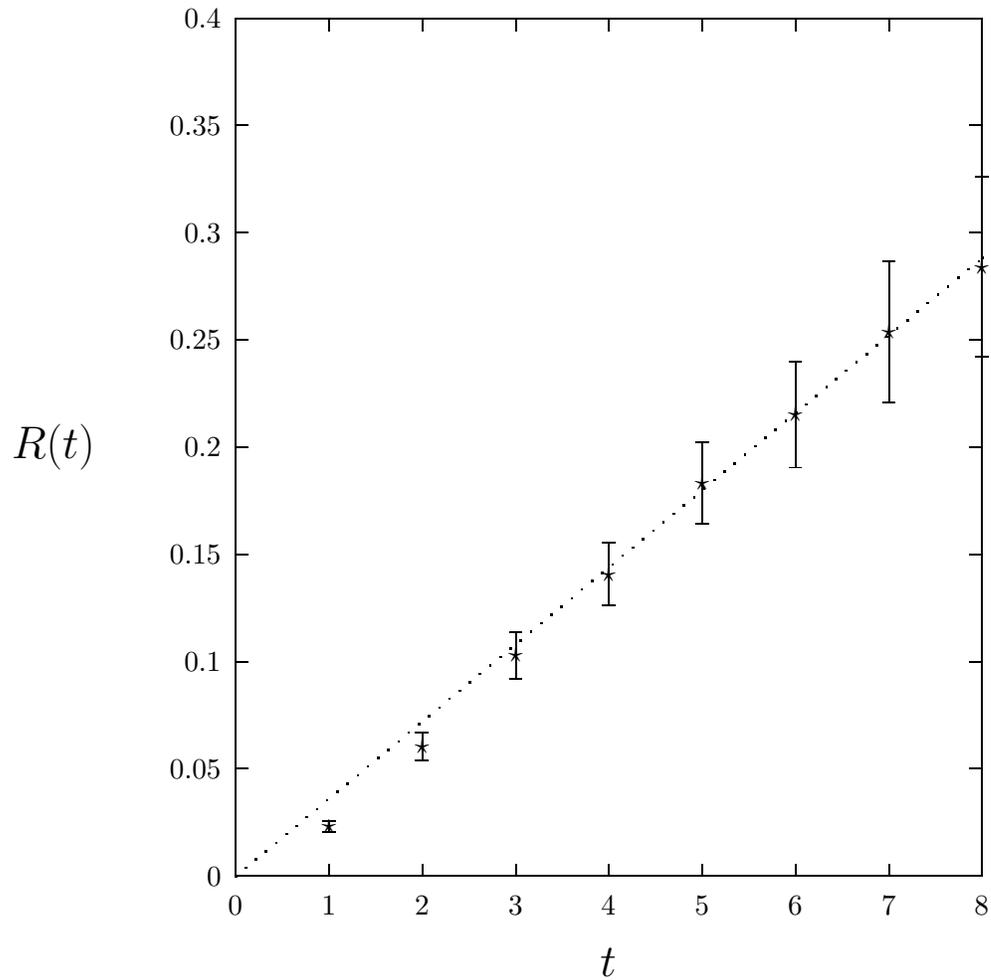
\begin{figure} 
\begin{center}
\setlength{\unitlength}{0.240900pt}
\ifx\plotpoint\undefined\newsavebox{\plotpoint}\fi
\sbox{\plotpoint}{\rule[-0.175pt]{0.350pt}{0.350pt}}%
\begin{picture}(1500,1619)(0,0)
\tenrm
\sbox{\plotpoint}{\rule[-0.175pt]{0.350pt}{0.350pt}}%
\put(264,158){\rule[-0.175pt]{282.335pt}{0.350pt}}
\put(264,158){\rule[-0.175pt]{4.818pt}{0.350pt}}
\put(242,158){\makebox(0,0)[r]{0}}
\put(1416,158){\rule[-0.175pt]{4.818pt}{0.350pt}}
\put(264,327){\rule[-0.175pt]{4.818pt}{0.350pt}}
\put(242,327){\makebox(0,0)[r]{0.05}}
\put(1416,327){\rule[-0.175pt]{4.818pt}{0.350pt}}
\put(264,495){\rule[-0.175pt]{4.818pt}{0.350pt}}
\put(242,495){\makebox(0,0)[r]{0.1}}
\put(1416,495){\rule[-0.175pt]{4.818pt}{0.350pt}}
\put(264,664){\rule[-0.175pt]{4.818pt}{0.350pt}}
\put(242,664){\makebox(0,0)[r]{0.15}}
\put(1416,664){\rule[-0.175pt]{4.818pt}{0.350pt}}
\put(264,832){\rule[-0.175pt]{4.818pt}{0.350pt}}
\put(242,832){\makebox(0,0)[r]{0.2}}
\put(1416,832){\rule[-0.175pt]{4.818pt}{0.350pt}}
\put(264,1001){\rule[-0.175pt]{4.818pt}{0.350pt}}
\put(242,1001){\makebox(0,0)[r]{0.25}}
\put(1416,1001){\rule[-0.175pt]{4.818pt}{0.350pt}}
\put(264,1169){\rule[-0.175pt]{4.818pt}{0.350pt}}
\put(242,1169){\makebox(0,0)[r]{0.3}}
\put(1416,1169){\rule[-0.175pt]{4.818pt}{0.350pt}}
\put(264,1338){\rule[-0.175pt]{4.818pt}{0.350pt}}
\put(242,1338){\makebox(0,0)[r]{0.35}}
\put(1416,1338){\rule[-0.175pt]{4.818pt}{0.350pt}}
\put(264,1506){\rule[-0.175pt]{4.818pt}{0.350pt}}
\put(242,1506){\makebox(0,0)[r]{0.4}}
\put(1416,1506){\rule[-0.175pt]{4.818pt}{0.350pt}}
\put(264,158){\rule[-0.175pt]{0.350pt}{4.818pt}}
\put(264,113){\makebox(0,0){0}}
\put(264,1486){\rule[-0.175pt]{0.350pt}{4.818pt}}
\put(411,158){\rule[-0.175pt]{0.350pt}{4.818pt}}
\put(411,113){\makebox(0,0){1}}
\put(411,1486){\rule[-0.175pt]{0.350pt}{4.818pt}}
\put(557,158){\rule[-0.175pt]{0.350pt}{4.818pt}}
\put(557,113){\makebox(0,0){2}}
\put(557,1486){\rule[-0.175pt]{0.350pt}{4.818pt}}
\put(704,158){\rule[-0.175pt]{0.350pt}{4.818pt}}
\put(704,113){\makebox(0,0){3}}
\put(704,1486){\rule[-0.175pt]{0.350pt}{4.818pt}}
\put(850,158){\rule[-0.175pt]{0.350pt}{4.818pt}}
\put(850,113){\makebox(0,0){4}}
\put(850,1486){\rule[-0.175pt]{0.350pt}{4.818pt}}
\put(997,158){\rule[-0.175pt]{0.350pt}{4.818pt}}
\put(997,113){\makebox(0,0){5}}
\put(997,1486){\rule[-0.175pt]{0.350pt}{4.818pt}}
\put(1143,158){\rule[-0.175pt]{0.350pt}{4.818pt}}
\put(1143,113){\makebox(0,0){6}}
\put(1143,1486){\rule[-0.175pt]{0.350pt}{4.818pt}}
\put(1290,158){\rule[-0.175pt]{0.350pt}{4.818pt}}
\put(1290,113){\makebox(0,0){7}}
\put(1290,1486){\rule[-0.175pt]{0.350pt}{4.818pt}}
\put(1436,158){\rule[-0.175pt]{0.350pt}{4.818pt}}
\put(1436,113){\makebox(0,0){8}}
\put(1436,1486){\rule[-0.175pt]{0.350pt}{4.818pt}}
\put(264,158){\rule[-0.175pt]{282.335pt}{0.350pt}}
\put(1436,158){\rule[-0.175pt]{0.350pt}{324.733pt}}
\put(264,1506){\rule[-0.175pt]{282.335pt}{0.350pt}}
\put(-87,832){\makebox(0,0)[l]{\shortstack{{\Large $R(t)$}}}}
\put(850,23){\makebox(0,0){{\Large $t$}}}
\put(264,158){\rule[-0.175pt]{0.350pt}{324.733pt}}
\put(411,236){\makebox(0,0){$\star$}}
\put(557,362){\makebox(0,0){$\star$}}
\put(704,505){\makebox(0,0){$\star$}}
\put(850,632){\makebox(0,0){$\star$}}
\put(997,776){\makebox(0,0){$\star$}}
\put(1143,883){\makebox(0,0){$\star$}}
\put(1290,1013){\makebox(0,0){$\star$}}
\put(1436,1115){\makebox(0,0){$\star$}}
\put(411,227){\rule[-0.175pt]{0.350pt}{4.336pt}}
\put(401,227){\rule[-0.175pt]{4.818pt}{0.350pt}}
\put(401,245){\rule[-0.175pt]{4.818pt}{0.350pt}}
\put(557,340){\rule[-0.175pt]{0.350pt}{10.600pt}}
\put(547,340){\rule[-0.175pt]{4.818pt}{0.350pt}}
\put(547,384){\rule[-0.175pt]{4.818pt}{0.350pt}}
\put(704,468){\rule[-0.175pt]{0.350pt}{17.827pt}}
\put(694,468){\rule[-0.175pt]{4.818pt}{0.350pt}}
\put(694,542){\rule[-0.175pt]{4.818pt}{0.350pt}}
\put(850,583){\rule[-0.175pt]{0.350pt}{23.849pt}}
\put(840,583){\rule[-0.175pt]{4.818pt}{0.350pt}}
\put(840,682){\rule[-0.175pt]{4.818pt}{0.350pt}}
\put(997,712){\rule[-0.175pt]{0.350pt}{30.835pt}}
\put(987,712){\rule[-0.175pt]{4.818pt}{0.350pt}}
\put(987,840){\rule[-0.175pt]{4.818pt}{0.350pt}}
\put(1143,800){\rule[-0.175pt]{0.350pt}{39.989pt}}
\put(1133,800){\rule[-0.175pt]{4.818pt}{0.350pt}}
\put(1133,966){\rule[-0.175pt]{4.818pt}{0.350pt}}
\put(1290,902){\rule[-0.175pt]{0.350pt}{53.480pt}}
\put(1280,902){\rule[-0.175pt]{4.818pt}{0.350pt}}
\put(1280,1124){\rule[-0.175pt]{4.818pt}{0.350pt}}
\put(1436,974){\rule[-0.175pt]{0.350pt}{68.175pt}}
\put(1426,974){\rule[-0.175pt]{4.818pt}{0.350pt}}
\put(1426,1257){\rule[-0.175pt]{4.818pt}{0.350pt}}
\sbox{\plotpoint}{\rule[-0.250pt]{0.500pt}{0.500pt}}%
\put(264,158){\usebox{\plotpoint}}
\put(264,158){\usebox{\plotpoint}}
\put(279,171){\usebox{\plotpoint}}
\put(295,184){\usebox{\plotpoint}}
\put(311,197){\usebox{\plotpoint}}
\put(327,210){\usebox{\plotpoint}}
\put(343,224){\usebox{\plotpoint}}
\put(359,237){\usebox{\plotpoint}}
\put(375,250){\usebox{\plotpoint}}
\put(391,263){\usebox{\plotpoint}}
\put(407,277){\usebox{\plotpoint}}
\put(423,290){\usebox{\plotpoint}}
\put(439,303){\usebox{\plotpoint}}
\put(455,316){\usebox{\plotpoint}}
\put(471,330){\usebox{\plotpoint}}
\put(487,343){\usebox{\plotpoint}}
\put(503,356){\usebox{\plotpoint}}
\put(519,369){\usebox{\plotpoint}}
\put(535,383){\usebox{\plotpoint}}
\put(551,396){\usebox{\plotpoint}}
\put(567,409){\usebox{\plotpoint}}
\put(583,422){\usebox{\plotpoint}}
\put(599,436){\usebox{\plotpoint}}
\put(615,449){\usebox{\plotpoint}}
\put(631,462){\usebox{\plotpoint}}
\put(647,475){\usebox{\plotpoint}}
\put(663,489){\usebox{\plotpoint}}
\put(679,502){\usebox{\plotpoint}}
\put(695,515){\usebox{\plotpoint}}
\put(711,528){\usebox{\plotpoint}}
\put(727,542){\usebox{\plotpoint}}
\put(743,555){\usebox{\plotpoint}}
\put(759,568){\usebox{\plotpoint}}
\put(775,581){\usebox{\plotpoint}}
\put(791,594){\usebox{\plotpoint}}
\put(807,608){\usebox{\plotpoint}}
\put(823,621){\usebox{\plotpoint}}
\put(839,634){\usebox{\plotpoint}}
\put(855,647){\usebox{\plotpoint}}
\put(871,661){\usebox{\plotpoint}}
\put(887,674){\usebox{\plotpoint}}
\put(903,687){\usebox{\plotpoint}}
\put(919,700){\usebox{\plotpoint}}
\put(935,714){\usebox{\plotpoint}}
\put(951,727){\usebox{\plotpoint}}
\put(967,740){\usebox{\plotpoint}}
\put(983,753){\usebox{\plotpoint}}
\put(999,767){\usebox{\plotpoint}}
\put(1015,780){\usebox{\plotpoint}}
\put(1031,793){\usebox{\plotpoint}}
\put(1047,806){\usebox{\plotpoint}}
\put(1063,820){\usebox{\plotpoint}}
\put(1079,833){\usebox{\plotpoint}}
\put(1095,846){\usebox{\plotpoint}}
\put(1111,859){\usebox{\plotpoint}}
\put(1127,873){\usebox{\plotpoint}}
\put(1143,886){\usebox{\plotpoint}}
\put(1159,899){\usebox{\plotpoint}}
\put(1175,912){\usebox{\plotpoint}}
\put(1191,926){\usebox{\plotpoint}}
\put(1206,939){\usebox{\plotpoint}}
\put(1222,952){\usebox{\plotpoint}}
\put(1238,965){\usebox{\plotpoint}}
\put(1254,978){\usebox{\plotpoint}}
\put(1270,992){\usebox{\plotpoint}}
\put(1286,1005){\usebox{\plotpoint}}
\put(1302,1018){\usebox{\plotpoint}}
\put(1318,1031){\usebox{\plotpoint}}
\put(1334,1045){\usebox{\plotpoint}}
\put(1350,1058){\usebox{\plotpoint}}
\put(1366,1071){\usebox{\plotpoint}}
\put(1382,1084){\usebox{\plotpoint}}
\put(1398,1098){\usebox{\plotpoint}}
\put(1414,1111){\usebox{\plotpoint}}
\put(1430,1124){\usebox{\plotpoint}}
\put(1436,1129){\usebox{\plotpoint}}
\end{picture}
\caption{The ratio $R(t)$ for $n_f = -4$; $R_2 = 0.5$.} 
\end{center}
\end{figure}

\begin{figure} 
\begin{center}
\setlength{\unitlength}{0.240900pt}
\ifx\plotpoint\undefined\newsavebox{\plotpoint}\fi
\sbox{\plotpoint}{\rule[-0.175pt]{0.350pt}{0.350pt}}%
\begin{picture}(1500,1619)(0,0)
\tenrm
\sbox{\plotpoint}{\rule[-0.175pt]{0.350pt}{0.350pt}}%
\put(264,158){\rule[-0.175pt]{282.335pt}{0.350pt}}
\put(264,158){\rule[-0.175pt]{4.818pt}{0.350pt}}
\put(242,158){\makebox(0,0)[r]{0}}
\put(1416,158){\rule[-0.175pt]{4.818pt}{0.350pt}}
\put(264,327){\rule[-0.175pt]{4.818pt}{0.350pt}}
\put(242,327){\makebox(0,0)[r]{0.05}}
\put(1416,327){\rule[-0.175pt]{4.818pt}{0.350pt}}
\put(264,495){\rule[-0.175pt]{4.818pt}{0.350pt}}
\put(242,495){\makebox(0,0)[r]{0.1}}
\put(1416,495){\rule[-0.175pt]{4.818pt}{0.350pt}}
\put(264,664){\rule[-0.175pt]{4.818pt}{0.350pt}}
\put(242,664){\makebox(0,0)[r]{0.15}}
\put(1416,664){\rule[-0.175pt]{4.818pt}{0.350pt}}
\put(264,832){\rule[-0.175pt]{4.818pt}{0.350pt}}
\put(242,832){\makebox(0,0)[r]{0.2}}
\put(1416,832){\rule[-0.175pt]{4.818pt}{0.350pt}}
\put(264,1001){\rule[-0.175pt]{4.818pt}{0.350pt}}
\put(242,1001){\makebox(0,0)[r]{0.25}}
\put(1416,1001){\rule[-0.175pt]{4.818pt}{0.350pt}}
\put(264,1169){\rule[-0.175pt]{4.818pt}{0.350pt}}
\put(242,1169){\makebox(0,0)[r]{0.3}}
\put(1416,1169){\rule[-0.175pt]{4.818pt}{0.350pt}}
\put(264,1338){\rule[-0.175pt]{4.818pt}{0.350pt}}
\put(242,1338){\makebox(0,0)[r]{0.35}}
\put(1416,1338){\rule[-0.175pt]{4.818pt}{0.350pt}}
\put(264,1506){\rule[-0.175pt]{4.818pt}{0.350pt}}
\put(242,1506){\makebox(0,0)[r]{0.4}}
\put(1416,1506){\rule[-0.175pt]{4.818pt}{0.350pt}}
\put(264,158){\rule[-0.175pt]{0.350pt}{4.818pt}}
\put(264,113){\makebox(0,0){0}}
\put(264,1486){\rule[-0.175pt]{0.350pt}{4.818pt}}
\put(411,158){\rule[-0.175pt]{0.350pt}{4.818pt}}
\put(411,113){\makebox(0,0){1}}
\put(411,1486){\rule[-0.175pt]{0.350pt}{4.818pt}}
\put(557,158){\rule[-0.175pt]{0.350pt}{4.818pt}}
\put(557,113){\makebox(0,0){2}}
\put(557,1486){\rule[-0.175pt]{0.350pt}{4.818pt}}
\put(704,158){\rule[-0.175pt]{0.350pt}{4.818pt}}
\put(704,113){\makebox(0,0){3}}
\put(704,1486){\rule[-0.175pt]{0.350pt}{4.818pt}}
\put(850,158){\rule[-0.175pt]{0.350pt}{4.818pt}}
\put(850,113){\makebox(0,0){4}}
\put(850,1486){\rule[-0.175pt]{0.350pt}{4.818pt}}
\put(997,158){\rule[-0.175pt]{0.350pt}{4.818pt}}
\put(997,113){\makebox(0,0){5}}
\put(997,1486){\rule[-0.175pt]{0.350pt}{4.818pt}}
\put(1143,158){\rule[-0.175pt]{0.350pt}{4.818pt}}
\put(1143,113){\makebox(0,0){6}}
\put(1143,1486){\rule[-0.175pt]{0.350pt}{4.818pt}}
\put(1290,158){\rule[-0.175pt]{0.350pt}{4.818pt}}
\put(1290,113){\makebox(0,0){7}}
\put(1290,1486){\rule[-0.175pt]{0.350pt}{4.818pt}}
\put(1436,158){\rule[-0.175pt]{0.350pt}{4.818pt}}
\put(1436,113){\makebox(0,0){8}}
\put(1436,1486){\rule[-0.175pt]{0.350pt}{4.818pt}}
\put(264,158){\rule[-0.175pt]{282.335pt}{0.350pt}}
\put(1436,158){\rule[-0.175pt]{0.350pt}{324.733pt}}
\put(264,1506){\rule[-0.175pt]{282.335pt}{0.350pt}}
\put(-87,832){\makebox(0,0)[l]{\shortstack{{\Large $R_0(t)$}}}}
\put(850,23){\makebox(0,0){{\Large $t$}}}
\put(264,158){\rule[-0.175pt]{0.350pt}{324.733pt}}
\put(264,171){\makebox(0,0){$\star$}}
\put(411,227){\makebox(0,0){$\star$}}
\put(557,329){\makebox(0,0){$\star$}}
\put(704,439){\makebox(0,0){$\star$}}
\put(850,572){\makebox(0,0){$\star$}}
\put(997,694){\makebox(0,0){$\star$}}
\put(1143,787){\makebox(0,0){$\star$}}
\put(1290,1029){\makebox(0,0){$\star$}}
\put(1436,1329){\makebox(0,0){$\star$}}
\put(264,168){\rule[-0.175pt]{0.350pt}{1.204pt}}
\put(254,168){\rule[-0.175pt]{4.818pt}{0.350pt}}
\put(254,173){\rule[-0.175pt]{4.818pt}{0.350pt}}
\put(411,213){\rule[-0.175pt]{0.350pt}{6.745pt}}
\put(401,213){\rule[-0.175pt]{4.818pt}{0.350pt}}
\put(401,241){\rule[-0.175pt]{4.818pt}{0.350pt}}
\put(557,292){\rule[-0.175pt]{0.350pt}{18.067pt}}
\put(547,292){\rule[-0.175pt]{4.818pt}{0.350pt}}
\put(547,367){\rule[-0.175pt]{4.818pt}{0.350pt}}
\put(704,370){\rule[-0.175pt]{0.350pt}{33.244pt}}
\put(694,370){\rule[-0.175pt]{4.818pt}{0.350pt}}
\put(694,508){\rule[-0.175pt]{4.818pt}{0.350pt}}
\put(850,455){\rule[-0.175pt]{0.350pt}{56.371pt}}
\put(840,455){\rule[-0.175pt]{4.818pt}{0.350pt}}
\put(840,689){\rule[-0.175pt]{4.818pt}{0.350pt}}
\put(997,518){\rule[-0.175pt]{0.350pt}{84.797pt}}
\put(987,518){\rule[-0.175pt]{4.818pt}{0.350pt}}
\put(987,870){\rule[-0.175pt]{4.818pt}{0.350pt}}
\put(1143,518){\rule[-0.175pt]{0.350pt}{129.363pt}}
\put(1133,518){\rule[-0.175pt]{4.818pt}{0.350pt}}
\put(1133,1055){\rule[-0.175pt]{4.818pt}{0.350pt}}
\put(1290,622){\rule[-0.175pt]{0.350pt}{196.093pt}}
\put(1280,622){\rule[-0.175pt]{4.818pt}{0.350pt}}
\put(1280,1436){\rule[-0.175pt]{4.818pt}{0.350pt}}
\put(1436,712){\rule[-0.175pt]{0.350pt}{191.275pt}}
\put(1426,712){\rule[-0.175pt]{4.818pt}{0.350pt}}
\put(1426,1506){\rule[-0.175pt]{4.818pt}{0.350pt}}
\sbox{\plotpoint}{\rule[-0.250pt]{0.500pt}{0.500pt}}%
\put(264,158){\usebox{\plotpoint}}
\put(264,158){\usebox{\plotpoint}}
\put(280,170){\usebox{\plotpoint}}
\put(297,182){\usebox{\plotpoint}}
\put(314,194){\usebox{\plotpoint}}
\put(331,206){\usebox{\plotpoint}}
\put(348,218){\usebox{\plotpoint}}
\put(365,230){\usebox{\plotpoint}}
\put(382,242){\usebox{\plotpoint}}
\put(399,254){\usebox{\plotpoint}}
\put(416,266){\usebox{\plotpoint}}
\put(432,278){\usebox{\plotpoint}}
\put(449,290){\usebox{\plotpoint}}
\put(466,302){\usebox{\plotpoint}}
\put(483,314){\usebox{\plotpoint}}
\put(500,326){\usebox{\plotpoint}}
\put(517,338){\usebox{\plotpoint}}
\put(534,350){\usebox{\plotpoint}}
\put(551,362){\usebox{\plotpoint}}
\put(568,374){\usebox{\plotpoint}}
\put(585,387){\usebox{\plotpoint}}
\put(601,399){\usebox{\plotpoint}}
\put(618,411){\usebox{\plotpoint}}
\put(635,423){\usebox{\plotpoint}}
\put(652,435){\usebox{\plotpoint}}
\put(669,447){\usebox{\plotpoint}}
\put(686,459){\usebox{\plotpoint}}
\put(703,471){\usebox{\plotpoint}}
\put(720,483){\usebox{\plotpoint}}
\put(737,495){\usebox{\plotpoint}}
\put(754,507){\usebox{\plotpoint}}
\put(770,519){\usebox{\plotpoint}}
\put(787,531){\usebox{\plotpoint}}
\put(804,543){\usebox{\plotpoint}}
\put(821,555){\usebox{\plotpoint}}
\put(838,567){\usebox{\plotpoint}}
\put(855,579){\usebox{\plotpoint}}
\put(872,591){\usebox{\plotpoint}}
\put(889,603){\usebox{\plotpoint}}
\put(906,616){\usebox{\plotpoint}}
\put(922,628){\usebox{\plotpoint}}
\put(939,640){\usebox{\plotpoint}}
\put(956,652){\usebox{\plotpoint}}
\put(973,664){\usebox{\plotpoint}}
\put(990,676){\usebox{\plotpoint}}
\put(1007,688){\usebox{\plotpoint}}
\put(1024,700){\usebox{\plotpoint}}
\put(1041,712){\usebox{\plotpoint}}
\put(1058,724){\usebox{\plotpoint}}
\put(1075,736){\usebox{\plotpoint}}
\put(1091,748){\usebox{\plotpoint}}
\put(1108,760){\usebox{\plotpoint}}
\put(1125,772){\usebox{\plotpoint}}
\put(1142,784){\usebox{\plotpoint}}
\put(1159,796){\usebox{\plotpoint}}
\put(1176,808){\usebox{\plotpoint}}
\put(1193,820){\usebox{\plotpoint}}
\put(1210,832){\usebox{\plotpoint}}
\put(1227,845){\usebox{\plotpoint}}
\put(1244,857){\usebox{\plotpoint}}
\put(1260,869){\usebox{\plotpoint}}
\put(1277,881){\usebox{\plotpoint}}
\put(1294,893){\usebox{\plotpoint}}
\put(1311,905){\usebox{\plotpoint}}
\put(1328,917){\usebox{\plotpoint}}
\put(1345,929){\usebox{\plotpoint}}
\put(1362,941){\usebox{\plotpoint}}
\put(1379,953){\usebox{\plotpoint}}
\put(1396,965){\usebox{\plotpoint}}
\put(1413,977){\usebox{\plotpoint}}
\put(1429,989){\usebox{\plotpoint}}
\put(1436,994){\usebox{\plotpoint}}
\end{picture}
\caption{The ratio $R_0(t)$ for $n_f = 0$; $R_2 = 0.5$.} 
\end{center}
\end{figure}
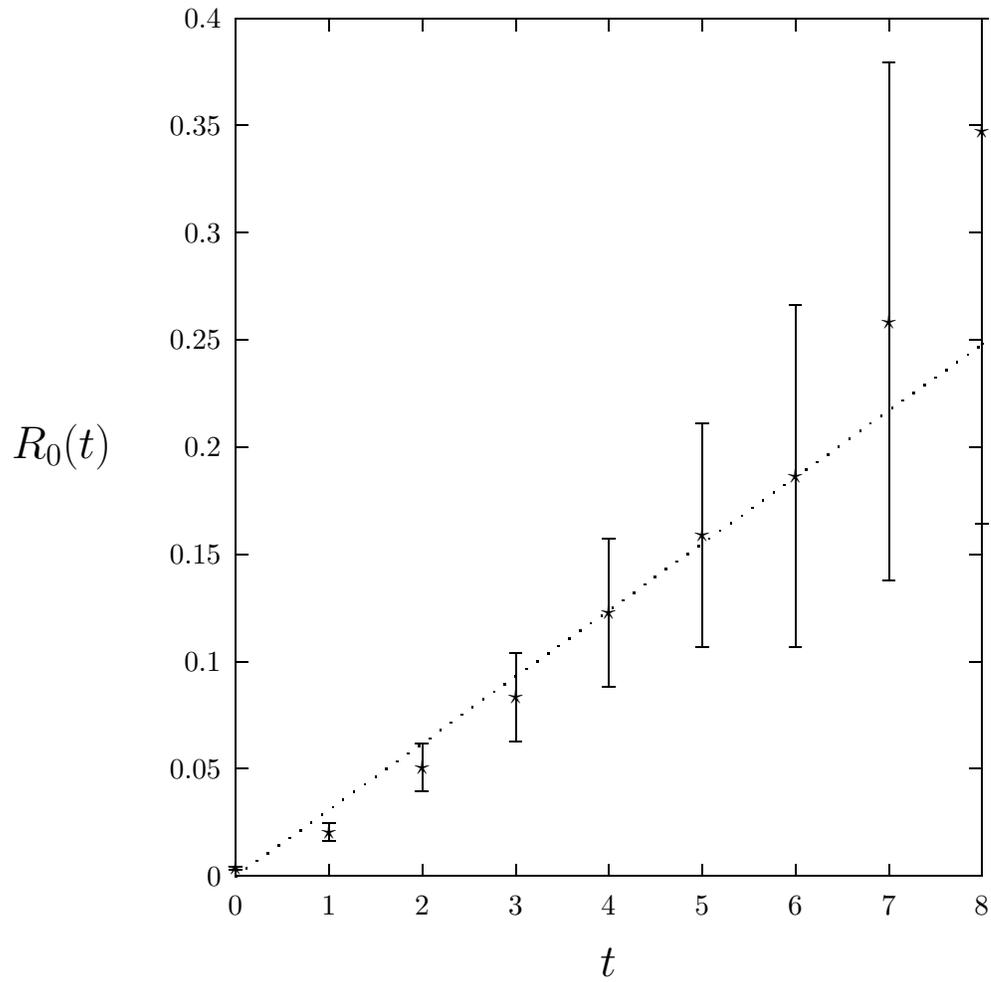

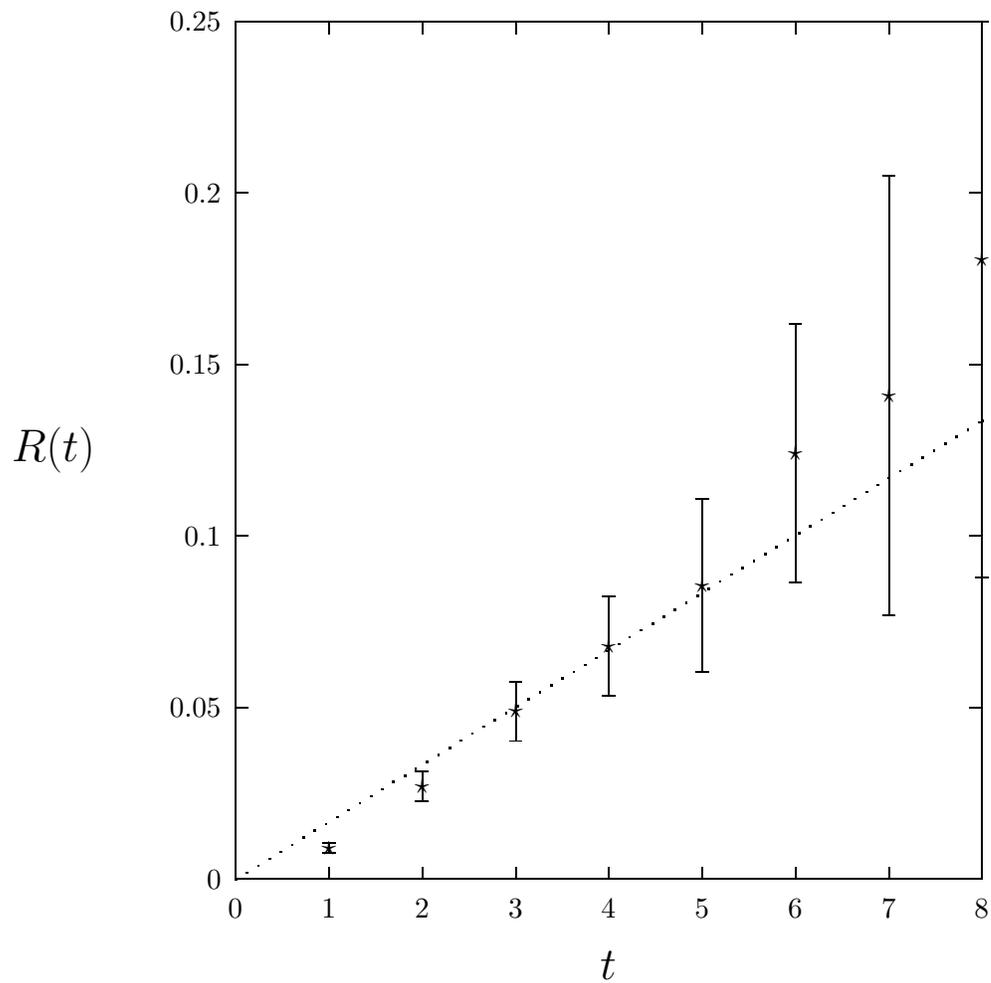
\begin{figure} 
\begin{center}
\setlength{\unitlength}{0.240900pt}
\ifx\plotpoint\undefined\newsavebox{\plotpoint}\fi
\sbox{\plotpoint}{\rule[-0.175pt]{0.350pt}{0.350pt}}%
\begin{picture}(1500,1619)(0,0)
\tenrm
\sbox{\plotpoint}{\rule[-0.175pt]{0.350pt}{0.350pt}}%
\put(264,158){\rule[-0.175pt]{282.335pt}{0.350pt}}
\put(264,158){\rule[-0.175pt]{4.818pt}{0.350pt}}
\put(242,158){\makebox(0,0)[r]{0}}
\put(1416,158){\rule[-0.175pt]{4.818pt}{0.350pt}}
\put(264,428){\rule[-0.175pt]{4.818pt}{0.350pt}}
\put(242,428){\makebox(0,0)[r]{0.05}}
\put(1416,428){\rule[-0.175pt]{4.818pt}{0.350pt}}
\put(264,697){\rule[-0.175pt]{4.818pt}{0.350pt}}
\put(242,697){\makebox(0,0)[r]{0.1}}
\put(1416,697){\rule[-0.175pt]{4.818pt}{0.350pt}}
\put(264,967){\rule[-0.175pt]{4.818pt}{0.350pt}}
\put(242,967){\makebox(0,0)[r]{0.15}}
\put(1416,967){\rule[-0.175pt]{4.818pt}{0.350pt}}
\put(264,1236){\rule[-0.175pt]{4.818pt}{0.350pt}}
\put(242,1236){\makebox(0,0)[r]{0.2}}
\put(1416,1236){\rule[-0.175pt]{4.818pt}{0.350pt}}
\put(264,1506){\rule[-0.175pt]{4.818pt}{0.350pt}}
\put(242,1506){\makebox(0,0)[r]{0.25}}
\put(1416,1506){\rule[-0.175pt]{4.818pt}{0.350pt}}
\put(264,158){\rule[-0.175pt]{0.350pt}{4.818pt}}
\put(264,113){\makebox(0,0){0}}
\put(264,1486){\rule[-0.175pt]{0.350pt}{4.818pt}}
\put(411,158){\rule[-0.175pt]{0.350pt}{4.818pt}}
\put(411,113){\makebox(0,0){1}}
\put(411,1486){\rule[-0.175pt]{0.350pt}{4.818pt}}
\put(557,158){\rule[-0.175pt]{0.350pt}{4.818pt}}
\put(557,113){\makebox(0,0){2}}
\put(557,1486){\rule[-0.175pt]{0.350pt}{4.818pt}}
\put(704,158){\rule[-0.175pt]{0.350pt}{4.818pt}}
\put(704,113){\makebox(0,0){3}}
\put(704,1486){\rule[-0.175pt]{0.350pt}{4.818pt}}
\put(850,158){\rule[-0.175pt]{0.350pt}{4.818pt}}
\put(850,113){\makebox(0,0){4}}
\put(850,1486){\rule[-0.175pt]{0.350pt}{4.818pt}}
\put(997,158){\rule[-0.175pt]{0.350pt}{4.818pt}}
\put(997,113){\makebox(0,0){5}}
\put(997,1486){\rule[-0.175pt]{0.350pt}{4.818pt}}
\put(1143,158){\rule[-0.175pt]{0.350pt}{4.818pt}}
\put(1143,113){\makebox(0,0){6}}
\put(1143,1486){\rule[-0.175pt]{0.350pt}{4.818pt}}
\put(1290,158){\rule[-0.175pt]{0.350pt}{4.818pt}}
\put(1290,113){\makebox(0,0){7}}
\put(1290,1486){\rule[-0.175pt]{0.350pt}{4.818pt}}
\put(1436,158){\rule[-0.175pt]{0.350pt}{4.818pt}}
\put(1436,113){\makebox(0,0){8}}
\put(1436,1486){\rule[-0.175pt]{0.350pt}{4.818pt}}
\put(264,158){\rule[-0.175pt]{282.335pt}{0.350pt}}
\put(1436,158){\rule[-0.175pt]{0.350pt}{324.733pt}}
\put(264,1506){\rule[-0.175pt]{282.335pt}{0.350pt}}
\put(-87,832){\makebox(0,0)[l]{\shortstack{{\Large $R(t)$}}}}
\put(850,23){\makebox(0,0){{\Large $t$}}}
\put(264,158){\rule[-0.175pt]{0.350pt}{324.733pt}}
\put(411,207){\makebox(0,0){$\star$}}
\put(557,304){\makebox(0,0){$\star$}}
\put(704,422){\makebox(0,0){$\star$}}
\put(850,524){\makebox(0,0){$\star$}}
\put(997,620){\makebox(0,0){$\star$}}
\put(1143,827){\makebox(0,0){$\star$}}
\put(1290,918){\makebox(0,0){$\star$}}
\put(1436,1132){\makebox(0,0){$\star$}}
\put(411,199){\rule[-0.175pt]{0.350pt}{3.854pt}}
\put(401,199){\rule[-0.175pt]{4.818pt}{0.350pt}}
\put(401,215){\rule[-0.175pt]{4.818pt}{0.350pt}}
\put(557,281){\rule[-0.175pt]{0.350pt}{11.322pt}}
\put(547,281){\rule[-0.175pt]{4.818pt}{0.350pt}}
\put(547,328){\rule[-0.175pt]{4.818pt}{0.350pt}}
\put(704,375){\rule[-0.175pt]{0.350pt}{22.404pt}}
\put(694,375){\rule[-0.175pt]{4.818pt}{0.350pt}}
\put(694,468){\rule[-0.175pt]{4.818pt}{0.350pt}}
\put(850,446){\rule[-0.175pt]{0.350pt}{37.580pt}}
\put(840,446){\rule[-0.175pt]{4.818pt}{0.350pt}}
\put(840,602){\rule[-0.175pt]{4.818pt}{0.350pt}}
\put(997,484){\rule[-0.175pt]{0.350pt}{65.525pt}}
\put(987,484){\rule[-0.175pt]{4.818pt}{0.350pt}}
\put(987,756){\rule[-0.175pt]{4.818pt}{0.350pt}}
\put(1143,624){\rule[-0.175pt]{0.350pt}{98.046pt}}
\put(1133,624){\rule[-0.175pt]{4.818pt}{0.350pt}}
\put(1133,1031){\rule[-0.175pt]{4.818pt}{0.350pt}}
\put(1290,572){\rule[-0.175pt]{0.350pt}{166.703pt}}
\put(1280,572){\rule[-0.175pt]{4.818pt}{0.350pt}}
\put(1280,1264){\rule[-0.175pt]{4.818pt}{0.350pt}}
\put(1436,632){\rule[-0.175pt]{0.350pt}{210.547pt}}
\put(1426,632){\rule[-0.175pt]{4.818pt}{0.350pt}}
\put(1426,1506){\rule[-0.175pt]{4.818pt}{0.350pt}}
\sbox{\plotpoint}{\rule[-0.250pt]{0.500pt}{0.500pt}}%
\put(264,158){\usebox{\plotpoint}}
\put(264,158){\usebox{\plotpoint}}
\put(281,168){\usebox{\plotpoint}}
\put(299,179){\usebox{\plotpoint}}
\put(317,190){\usebox{\plotpoint}}
\put(334,201){\usebox{\plotpoint}}
\put(352,212){\usebox{\plotpoint}}
\put(370,223){\usebox{\plotpoint}}
\put(387,234){\usebox{\plotpoint}}
\put(405,244){\usebox{\plotpoint}}
\put(423,255){\usebox{\plotpoint}}
\put(440,266){\usebox{\plotpoint}}
\put(458,277){\usebox{\plotpoint}}
\put(476,288){\usebox{\plotpoint}}
\put(493,299){\usebox{\plotpoint}}
\put(511,310){\usebox{\plotpoint}}
\put(529,320){\usebox{\plotpoint}}
\put(546,331){\usebox{\plotpoint}}
\put(564,342){\usebox{\plotpoint}}
\put(582,353){\usebox{\plotpoint}}
\put(600,364){\usebox{\plotpoint}}
\put(617,375){\usebox{\plotpoint}}
\put(635,386){\usebox{\plotpoint}}
\put(653,397){\usebox{\plotpoint}}
\put(670,407){\usebox{\plotpoint}}
\put(688,418){\usebox{\plotpoint}}
\put(706,429){\usebox{\plotpoint}}
\put(723,440){\usebox{\plotpoint}}
\put(741,451){\usebox{\plotpoint}}
\put(759,462){\usebox{\plotpoint}}
\put(776,473){\usebox{\plotpoint}}
\put(794,483){\usebox{\plotpoint}}
\put(812,494){\usebox{\plotpoint}}
\put(829,505){\usebox{\plotpoint}}
\put(847,516){\usebox{\plotpoint}}
\put(865,527){\usebox{\plotpoint}}
\put(882,538){\usebox{\plotpoint}}
\put(900,549){\usebox{\plotpoint}}
\put(918,559){\usebox{\plotpoint}}
\put(936,570){\usebox{\plotpoint}}
\put(953,581){\usebox{\plotpoint}}
\put(971,592){\usebox{\plotpoint}}
\put(989,603){\usebox{\plotpoint}}
\put(1006,614){\usebox{\plotpoint}}
\put(1024,625){\usebox{\plotpoint}}
\put(1042,636){\usebox{\plotpoint}}
\put(1059,646){\usebox{\plotpoint}}
\put(1077,657){\usebox{\plotpoint}}
\put(1095,668){\usebox{\plotpoint}}
\put(1112,679){\usebox{\plotpoint}}
\put(1130,690){\usebox{\plotpoint}}
\put(1148,701){\usebox{\plotpoint}}
\put(1165,712){\usebox{\plotpoint}}
\put(1183,722){\usebox{\plotpoint}}
\put(1201,733){\usebox{\plotpoint}}
\put(1218,744){\usebox{\plotpoint}}
\put(1236,755){\usebox{\plotpoint}}
\put(1254,766){\usebox{\plotpoint}}
\put(1272,777){\usebox{\plotpoint}}
\put(1289,788){\usebox{\plotpoint}}
\put(1307,799){\usebox{\plotpoint}}
\put(1325,809){\usebox{\plotpoint}}
\put(1342,820){\usebox{\plotpoint}}
\put(1360,831){\usebox{\plotpoint}}
\put(1378,842){\usebox{\plotpoint}}
\put(1395,853){\usebox{\plotpoint}}
\put(1413,864){\usebox{\plotpoint}}
\put(1431,875){\usebox{\plotpoint}}
\put(1436,878){\usebox{\plotpoint}}
\end{picture}
\caption{The ratio $R(t)$ for $n_f = -4$; $R_2 = 0.6$.} 
\end{center}
\end{figure}

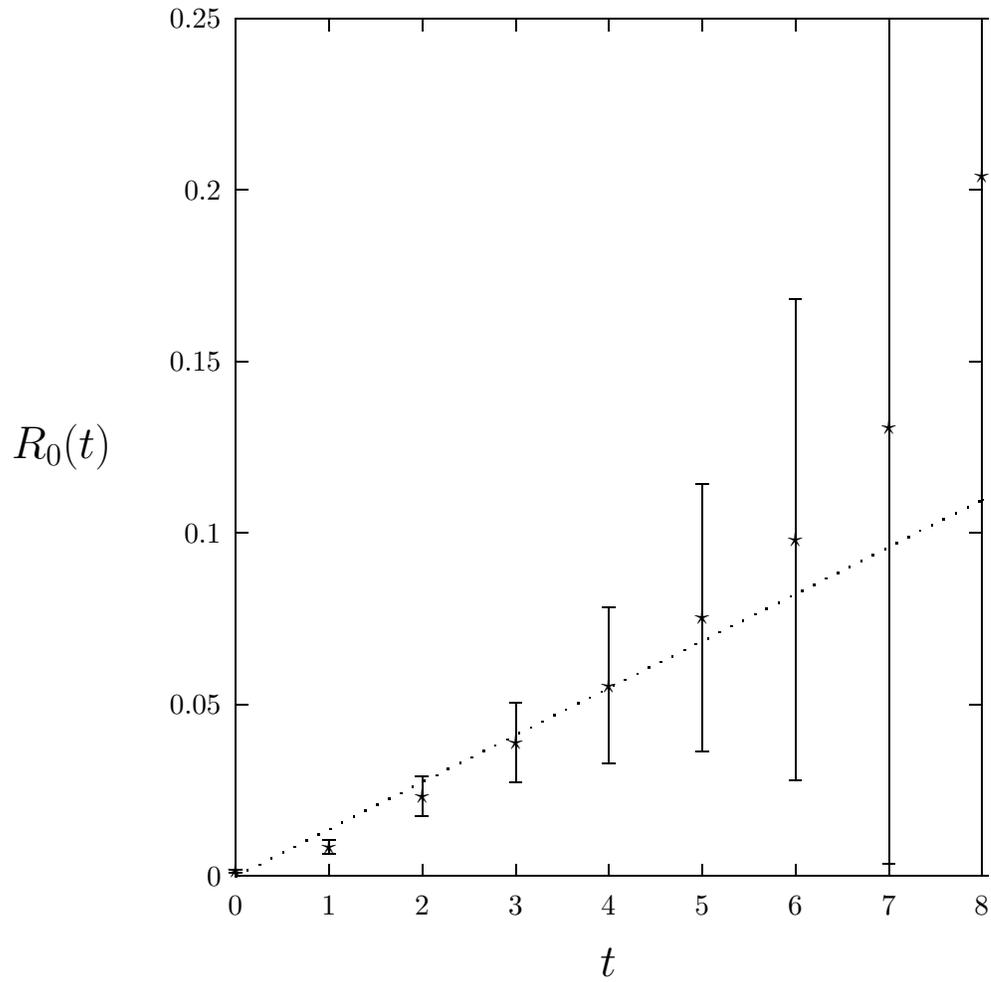
\begin{figure} 
\begin{center}
\setlength{\unitlength}{0.240900pt}
\ifx\plotpoint\undefined\newsavebox{\plotpoint}\fi
\begin{picture}(1500,1619)(0,0)
\tenrm
\sbox{\plotpoint}{\rule[-0.175pt]{0.350pt}{0.350pt}}%
\put(264,158){\rule[-0.175pt]{282.335pt}{0.350pt}}
\put(264,158){\rule[-0.175pt]{4.818pt}{0.350pt}}
\put(242,158){\makebox(0,0)[r]{0}}
\put(1416,158){\rule[-0.175pt]{4.818pt}{0.350pt}}
\put(264,428){\rule[-0.175pt]{4.818pt}{0.350pt}}
\put(242,428){\makebox(0,0)[r]{0.05}}
\put(1416,428){\rule[-0.175pt]{4.818pt}{0.350pt}}
\put(264,697){\rule[-0.175pt]{4.818pt}{0.350pt}}
\put(242,697){\makebox(0,0)[r]{0.1}}
\put(1416,697){\rule[-0.175pt]{4.818pt}{0.350pt}}
\put(264,967){\rule[-0.175pt]{4.818pt}{0.350pt}}
\put(242,967){\makebox(0,0)[r]{0.15}}
\put(1416,967){\rule[-0.175pt]{4.818pt}{0.350pt}}
\put(264,1236){\rule[-0.175pt]{4.818pt}{0.350pt}}
\put(242,1236){\makebox(0,0)[r]{0.2}}
\put(1416,1236){\rule[-0.175pt]{4.818pt}{0.350pt}}
\put(264,1506){\rule[-0.175pt]{4.818pt}{0.350pt}}
\put(242,1506){\makebox(0,0)[r]{0.25}}
\put(1416,1506){\rule[-0.175pt]{4.818pt}{0.350pt}}
\put(264,158){\rule[-0.175pt]{0.350pt}{4.818pt}}
\put(264,113){\makebox(0,0){0}}
\put(264,1486){\rule[-0.175pt]{0.350pt}{4.818pt}}
\put(411,158){\rule[-0.175pt]{0.350pt}{4.818pt}}
\put(411,113){\makebox(0,0){1}}
\put(411,1486){\rule[-0.175pt]{0.350pt}{4.818pt}}
\put(557,158){\rule[-0.175pt]{0.350pt}{4.818pt}}
\put(557,113){\makebox(0,0){2}}
\put(557,1486){\rule[-0.175pt]{0.350pt}{4.818pt}}
\put(704,158){\rule[-0.175pt]{0.350pt}{4.818pt}}
\put(704,113){\makebox(0,0){3}}
\put(704,1486){\rule[-0.175pt]{0.350pt}{4.818pt}}
\put(850,158){\rule[-0.175pt]{0.350pt}{4.818pt}}
\put(850,113){\makebox(0,0){4}}
\put(850,1486){\rule[-0.175pt]{0.350pt}{4.818pt}}
\put(997,158){\rule[-0.175pt]{0.350pt}{4.818pt}}
\put(997,113){\makebox(0,0){5}}
\put(997,1486){\rule[-0.175pt]{0.350pt}{4.818pt}}
\put(1143,158){\rule[-0.175pt]{0.350pt}{4.818pt}}
\put(1143,113){\makebox(0,0){6}}
\put(1143,1486){\rule[-0.175pt]{0.350pt}{4.818pt}}
\put(1290,158){\rule[-0.175pt]{0.350pt}{4.818pt}}
\put(1290,113){\makebox(0,0){7}}
\put(1290,1486){\rule[-0.175pt]{0.350pt}{4.818pt}}
\put(1436,158){\rule[-0.175pt]{0.350pt}{4.818pt}}
\put(1436,113){\makebox(0,0){8}}
\put(1436,1486){\rule[-0.175pt]{0.350pt}{4.818pt}}
\put(264,158){\rule[-0.175pt]{282.335pt}{0.350pt}}
\put(1436,158){\rule[-0.175pt]{0.350pt}{324.733pt}}
\put(264,1506){\rule[-0.175pt]{282.335pt}{0.350pt}}
\put(-87,832){\makebox(0,0)[l]{\shortstack{{\Large $R_0(t)$}}}}
\put(850,23){\makebox(0,0){{\Large $t$}}}
\put(264,158){\rule[-0.175pt]{0.350pt}{324.733pt}}
\put(264,166){\makebox(0,0){$\star$}}
\put(411,204){\makebox(0,0){$\star$}}
\put(557,284){\makebox(0,0){$\star$}}
\put(704,368){\makebox(0,0){$\star$}}
\put(850,457){\makebox(0,0){$\star$}}
\put(997,564){\makebox(0,0){$\star$}}
\put(1143,687){\makebox(0,0){$\star$}}
\put(1290,863){\makebox(0,0){$\star$}}
\put(1436,1258){\makebox(0,0){$\star$}}
\put(264,164){\rule[-0.175pt]{0.350pt}{0.964pt}}
\put(254,164){\rule[-0.175pt]{4.818pt}{0.350pt}}
\put(254,168){\rule[-0.175pt]{4.818pt}{0.350pt}}
\put(411,193){\rule[-0.175pt]{0.350pt}{5.300pt}}
\put(401,193){\rule[-0.175pt]{4.818pt}{0.350pt}}
\put(401,215){\rule[-0.175pt]{4.818pt}{0.350pt}}
\put(557,253){\rule[-0.175pt]{0.350pt}{14.936pt}}
\put(547,253){\rule[-0.175pt]{4.818pt}{0.350pt}}
\put(547,315){\rule[-0.175pt]{4.818pt}{0.350pt}}
\put(704,306){\rule[-0.175pt]{0.350pt}{29.872pt}}
\put(694,306){\rule[-0.175pt]{4.818pt}{0.350pt}}
\put(694,430){\rule[-0.175pt]{4.818pt}{0.350pt}}
\put(850,335){\rule[-0.175pt]{0.350pt}{59.020pt}}
\put(840,335){\rule[-0.175pt]{4.818pt}{0.350pt}}
\put(840,580){\rule[-0.175pt]{4.818pt}{0.350pt}}
\put(997,354){\rule[-0.175pt]{0.350pt}{101.178pt}}
\put(987,354){\rule[-0.175pt]{4.818pt}{0.350pt}}
\put(987,774){\rule[-0.175pt]{4.818pt}{0.350pt}}
\put(1143,309){\rule[-0.175pt]{0.350pt}{182.120pt}}
\put(1133,309){\rule[-0.175pt]{4.818pt}{0.350pt}}
\put(1133,1065){\rule[-0.175pt]{4.818pt}{0.350pt}}
\put(1290,177){\rule[-0.175pt]{0.350pt}{320.156pt}}
\put(1280,177){\rule[-0.175pt]{4.818pt}{0.350pt}}
\put(1280,1506){\rule[-0.175pt]{4.818pt}{0.350pt}}
\put(1436,158){\rule[-0.175pt]{0.350pt}{324.733pt}}
\put(1426,158){\rule[-0.175pt]{4.818pt}{0.350pt}}
\put(1426,1506){\rule[-0.175pt]{4.818pt}{0.350pt}}
\sbox{\plotpoint}{\rule[-0.250pt]{0.500pt}{0.500pt}}%
\put(264,158){\usebox{\plotpoint}}
\put(264,158){\usebox{\plotpoint}}
\put(282,167){\usebox{\plotpoint}}
\put(301,176){\usebox{\plotpoint}}
\put(319,186){\usebox{\plotpoint}}
\put(338,195){\usebox{\plotpoint}}
\put(356,204){\usebox{\plotpoint}}
\put(375,214){\usebox{\plotpoint}}
\put(393,223){\usebox{\plotpoint}}
\put(412,232){\usebox{\plotpoint}}
\put(430,242){\usebox{\plotpoint}}
\put(449,251){\usebox{\plotpoint}}
\put(467,260){\usebox{\plotpoint}}
\put(486,270){\usebox{\plotpoint}}
\put(504,279){\usebox{\plotpoint}}
\put(523,288){\usebox{\plotpoint}}
\put(541,298){\usebox{\plotpoint}}
\put(560,307){\usebox{\plotpoint}}
\put(579,316){\usebox{\plotpoint}}
\put(597,326){\usebox{\plotpoint}}
\put(616,335){\usebox{\plotpoint}}
\put(634,344){\usebox{\plotpoint}}
\put(653,354){\usebox{\plotpoint}}
\put(671,363){\usebox{\plotpoint}}
\put(690,372){\usebox{\plotpoint}}
\put(708,382){\usebox{\plotpoint}}
\put(727,391){\usebox{\plotpoint}}
\put(745,400){\usebox{\plotpoint}}
\put(764,410){\usebox{\plotpoint}}
\put(782,419){\usebox{\plotpoint}}
\put(801,429){\usebox{\plotpoint}}
\put(819,438){\usebox{\plotpoint}}
\put(838,447){\usebox{\plotpoint}}
\put(857,457){\usebox{\plotpoint}}
\put(875,466){\usebox{\plotpoint}}
\put(894,475){\usebox{\plotpoint}}
\put(912,485){\usebox{\plotpoint}}
\put(931,494){\usebox{\plotpoint}}
\put(949,503){\usebox{\plotpoint}}
\put(968,513){\usebox{\plotpoint}}
\put(986,522){\usebox{\plotpoint}}
\put(1005,531){\usebox{\plotpoint}}
\put(1023,541){\usebox{\plotpoint}}
\put(1042,550){\usebox{\plotpoint}}
\put(1060,559){\usebox{\plotpoint}}
\put(1079,569){\usebox{\plotpoint}}
\put(1097,578){\usebox{\plotpoint}}
\put(1116,587){\usebox{\plotpoint}}
\put(1135,597){\usebox{\plotpoint}}
\put(1153,606){\usebox{\plotpoint}}
\put(1172,615){\usebox{\plotpoint}}
\put(1190,625){\usebox{\plotpoint}}
\put(1209,634){\usebox{\plotpoint}}
\put(1227,643){\usebox{\plotpoint}}
\put(1246,653){\usebox{\plotpoint}}
\put(1264,662){\usebox{\plotpoint}}
\put(1283,671){\usebox{\plotpoint}}
\put(1301,681){\usebox{\plotpoint}}
\put(1320,690){\usebox{\plotpoint}}
\put(1338,700){\usebox{\plotpoint}}
\put(1357,709){\usebox{\plotpoint}}
\put(1375,718){\usebox{\plotpoint}}
\put(1394,728){\usebox{\plotpoint}}
\put(1413,737){\usebox{\plotpoint}}
\put(1431,746){\usebox{\plotpoint}}
\put(1436,749){\usebox{\plotpoint}}
\end{picture}
\caption{The ratio $R_0(t)$ for $n_f = 0$; $R_2 = 0.6$.} 
\end{center}
\end{figure}

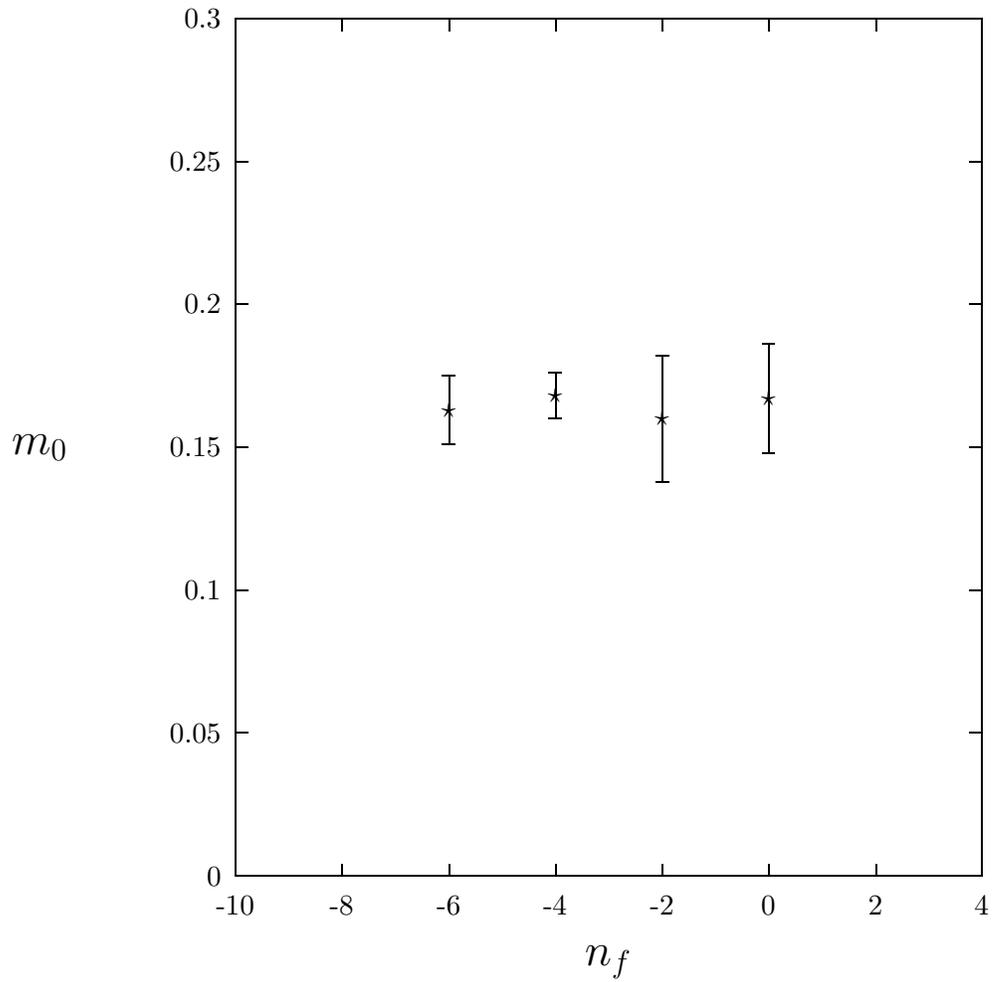
\begin{figure} 
\begin{center}
\setlength{\unitlength}{0.240900pt}
\ifx\plotpoint\undefined\newsavebox{\plotpoint}\fi
\sbox{\plotpoint}{\rule[-0.175pt]{0.350pt}{0.350pt}}%
\begin{picture}(1500,1619)(0,0)
\tenrm
\sbox{\plotpoint}{\rule[-0.175pt]{0.350pt}{0.350pt}}%
\put(264,158){\rule[-0.175pt]{282.335pt}{0.350pt}}
\put(264,158){\rule[-0.175pt]{4.818pt}{0.350pt}}
\put(242,158){\makebox(0,0)[r]{0}}
\put(1416,158){\rule[-0.175pt]{4.818pt}{0.350pt}}
\put(264,383){\rule[-0.175pt]{4.818pt}{0.350pt}}
\put(242,383){\makebox(0,0)[r]{0.05}}
\put(1416,383){\rule[-0.175pt]{4.818pt}{0.350pt}}
\put(264,607){\rule[-0.175pt]{4.818pt}{0.350pt}}
\put(242,607){\makebox(0,0)[r]{0.1}}
\put(1416,607){\rule[-0.175pt]{4.818pt}{0.350pt}}
\put(264,832){\rule[-0.175pt]{4.818pt}{0.350pt}}
\put(242,832){\makebox(0,0)[r]{0.15}}
\put(1416,832){\rule[-0.175pt]{4.818pt}{0.350pt}}
\put(264,1057){\rule[-0.175pt]{4.818pt}{0.350pt}}
\put(242,1057){\makebox(0,0)[r]{0.2}}
\put(1416,1057){\rule[-0.175pt]{4.818pt}{0.350pt}}
\put(264,1281){\rule[-0.175pt]{4.818pt}{0.350pt}}
\put(242,1281){\makebox(0,0)[r]{0.25}}
\put(1416,1281){\rule[-0.175pt]{4.818pt}{0.350pt}}
\put(264,1506){\rule[-0.175pt]{4.818pt}{0.350pt}}
\put(242,1506){\makebox(0,0)[r]{0.3}}
\put(1416,1506){\rule[-0.175pt]{4.818pt}{0.350pt}}
\put(264,158){\rule[-0.175pt]{0.350pt}{4.818pt}}
\put(264,113){\makebox(0,0){-10}}
\put(264,1486){\rule[-0.175pt]{0.350pt}{4.818pt}}
\put(431,158){\rule[-0.175pt]{0.350pt}{4.818pt}}
\put(431,113){\makebox(0,0){-8}}
\put(431,1486){\rule[-0.175pt]{0.350pt}{4.818pt}}
\put(599,158){\rule[-0.175pt]{0.350pt}{4.818pt}}
\put(599,113){\makebox(0,0){-6}}
\put(599,1486){\rule[-0.175pt]{0.350pt}{4.818pt}}
\put(766,158){\rule[-0.175pt]{0.350pt}{4.818pt}}
\put(766,113){\makebox(0,0){-4}}
\put(766,1486){\rule[-0.175pt]{0.350pt}{4.818pt}}
\put(934,158){\rule[-0.175pt]{0.350pt}{4.818pt}}
\put(934,113){\makebox(0,0){-2}}
\put(934,1486){\rule[-0.175pt]{0.350pt}{4.818pt}}
\put(1101,158){\rule[-0.175pt]{0.350pt}{4.818pt}}
\put(1101,113){\makebox(0,0){0}}
\put(1101,1486){\rule[-0.175pt]{0.350pt}{4.818pt}}
\put(1269,158){\rule[-0.175pt]{0.350pt}{4.818pt}}
\put(1269,113){\makebox(0,0){2}}
\put(1269,1486){\rule[-0.175pt]{0.350pt}{4.818pt}}
\put(1436,158){\rule[-0.175pt]{0.350pt}{4.818pt}}
\put(1436,113){\makebox(0,0){4}}
\put(1436,1486){\rule[-0.175pt]{0.350pt}{4.818pt}}
\put(264,158){\rule[-0.175pt]{282.335pt}{0.350pt}}
\put(1436,158){\rule[-0.175pt]{0.350pt}{324.733pt}}
\put(264,1506){\rule[-0.175pt]{282.335pt}{0.350pt}}
\put(-87,832){\makebox(0,0)[l]{\shortstack{{\Large $m_0$}}}}
\put(850,23){\makebox(0,0){{\Large $n_f$}}}
\put(264,158){\rule[-0.175pt]{0.350pt}{324.733pt}}
\put(599,890){\makebox(0,0){$\star$}}
\put(766,913){\makebox(0,0){$\star$}}
\put(934,877){\makebox(0,0){$\star$}}
\put(1101,908){\makebox(0,0){$\star$}}
\put(599,836){\rule[-0.175pt]{0.350pt}{26.017pt}}
\put(589,836){\rule[-0.175pt]{4.818pt}{0.350pt}}
\put(589,944){\rule[-0.175pt]{4.818pt}{0.350pt}}
\put(766,877){\rule[-0.175pt]{0.350pt}{17.345pt}}
\put(756,877){\rule[-0.175pt]{4.818pt}{0.350pt}}
\put(756,949){\rule[-0.175pt]{4.818pt}{0.350pt}}
\put(934,778){\rule[-0.175pt]{0.350pt}{47.698pt}}
\put(924,778){\rule[-0.175pt]{4.818pt}{0.350pt}}
\put(924,976){\rule[-0.175pt]{4.818pt}{0.350pt}}
\put(1101,823){\rule[-0.175pt]{0.350pt}{41.194pt}}
\put(1091,823){\rule[-0.175pt]{4.818pt}{0.350pt}}
\put(1091,994){\rule[-0.175pt]{4.818pt}{0.350pt}}
\end{picture}
\caption{$m_0$ as a function of $n_f$ for $R_2 = 0.5$.} 
\end{center}
\end{figure}

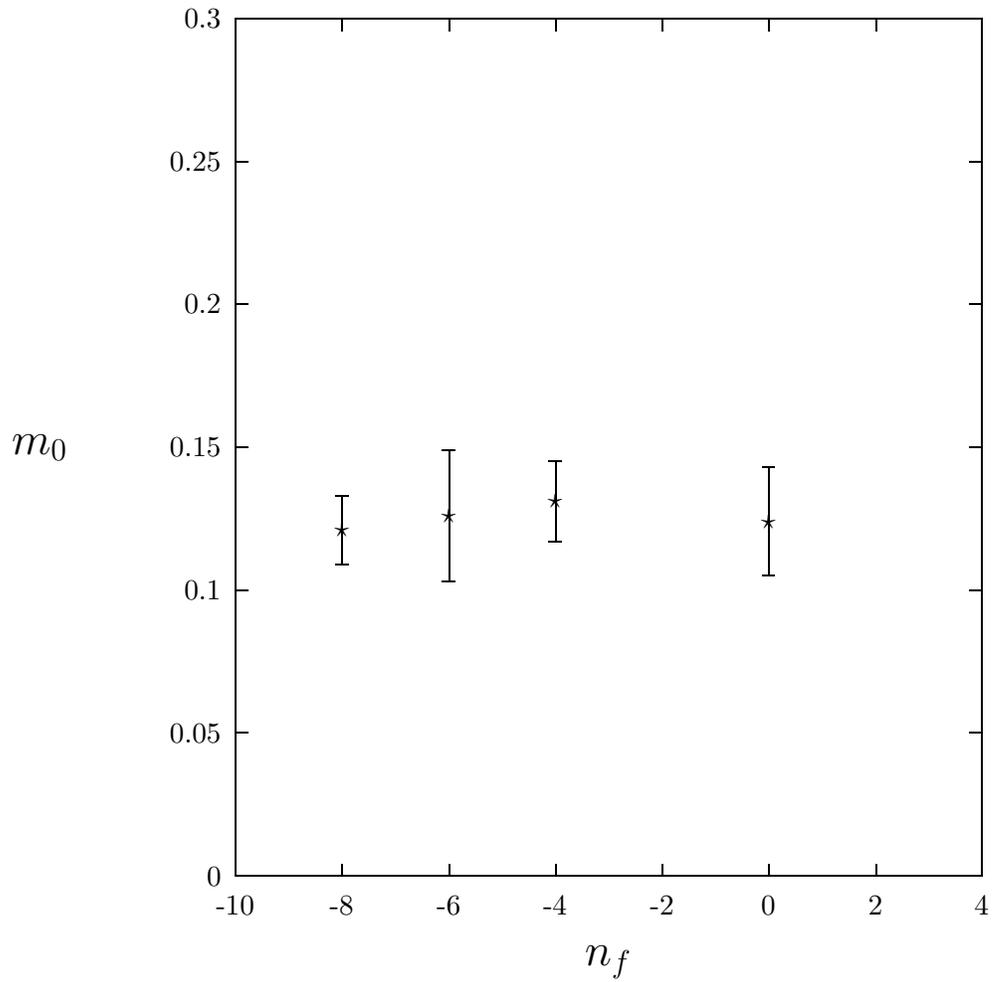
\begin{figure} 
\begin{center}
\setlength{\unitlength}{0.240900pt}
\ifx\plotpoint\undefined\newsavebox{\plotpoint}\fi
\begin{picture}(1500,1619)(0,0)
\tenrm
\sbox{\plotpoint}{\rule[-0.175pt]{0.350pt}{0.350pt}}%
\put(264,158){\rule[-0.175pt]{282.335pt}{0.350pt}}
\put(264,158){\rule[-0.175pt]{4.818pt}{0.350pt}}
\put(242,158){\makebox(0,0)[r]{0}}
\put(1416,158){\rule[-0.175pt]{4.818pt}{0.350pt}}
\put(264,383){\rule[-0.175pt]{4.818pt}{0.350pt}}
\put(242,383){\makebox(0,0)[r]{0.05}}
\put(1416,383){\rule[-0.175pt]{4.818pt}{0.350pt}}
\put(264,607){\rule[-0.175pt]{4.818pt}{0.350pt}}
\put(242,607){\makebox(0,0)[r]{0.1}}
\put(1416,607){\rule[-0.175pt]{4.818pt}{0.350pt}}
\put(264,832){\rule[-0.175pt]{4.818pt}{0.350pt}}
\put(242,832){\makebox(0,0)[r]{0.15}}
\put(1416,832){\rule[-0.175pt]{4.818pt}{0.350pt}}
\put(264,1057){\rule[-0.175pt]{4.818pt}{0.350pt}}
\put(242,1057){\makebox(0,0)[r]{0.2}}
\put(1416,1057){\rule[-0.175pt]{4.818pt}{0.350pt}}
\put(264,1281){\rule[-0.175pt]{4.818pt}{0.350pt}}
\put(242,1281){\makebox(0,0)[r]{0.25}}
\put(1416,1281){\rule[-0.175pt]{4.818pt}{0.350pt}}
\put(264,1506){\rule[-0.175pt]{4.818pt}{0.350pt}}
\put(242,1506){\makebox(0,0)[r]{0.3}}
\put(1416,1506){\rule[-0.175pt]{4.818pt}{0.350pt}}
\put(264,158){\rule[-0.175pt]{0.350pt}{4.818pt}}
\put(264,113){\makebox(0,0){-10}}
\put(264,1486){\rule[-0.175pt]{0.350pt}{4.818pt}}
\put(431,158){\rule[-0.175pt]{0.350pt}{4.818pt}}
\put(431,113){\makebox(0,0){-8}}
\put(431,1486){\rule[-0.175pt]{0.350pt}{4.818pt}}
\put(599,158){\rule[-0.175pt]{0.350pt}{4.818pt}}
\put(599,113){\makebox(0,0){-6}}
\put(599,1486){\rule[-0.175pt]{0.350pt}{4.818pt}}
\put(766,158){\rule[-0.175pt]{0.350pt}{4.818pt}}
\put(766,113){\makebox(0,0){-4}}
\put(766,1486){\rule[-0.175pt]{0.350pt}{4.818pt}}
\put(934,158){\rule[-0.175pt]{0.350pt}{4.818pt}}
\put(934,113){\makebox(0,0){-2}}
\put(934,1486){\rule[-0.175pt]{0.350pt}{4.818pt}}
\put(1101,158){\rule[-0.175pt]{0.350pt}{4.818pt}}
\put(1101,113){\makebox(0,0){0}}
\put(1101,1486){\rule[-0.175pt]{0.350pt}{4.818pt}}
\put(1269,158){\rule[-0.175pt]{0.350pt}{4.818pt}}
\put(1269,113){\makebox(0,0){2}}
\put(1269,1486){\rule[-0.175pt]{0.350pt}{4.818pt}}
\put(1436,158){\rule[-0.175pt]{0.350pt}{4.818pt}}
\put(1436,113){\makebox(0,0){4}}
\put(1436,1486){\rule[-0.175pt]{0.350pt}{4.818pt}}
\put(264,158){\rule[-0.175pt]{282.335pt}{0.350pt}}
\put(1436,158){\rule[-0.175pt]{0.350pt}{324.733pt}}
\put(264,1506){\rule[-0.175pt]{282.335pt}{0.350pt}}
\put(-87,832){\makebox(0,0)[l]{\shortstack{{\Large $m_0$}}}}
\put(850,23){\makebox(0,0){{\Large $n_f$}}}
\put(264,158){\rule[-0.175pt]{0.350pt}{324.733pt}}
\put(431,702){\makebox(0,0){$\star$}}
\put(599,724){\makebox(0,0){$\star$}}
\put(766,747){\makebox(0,0){$\star$}}
\put(1101,715){\makebox(0,0){$\star$}}
\put(431,648){\rule[-0.175pt]{0.350pt}{26.017pt}}
\put(421,648){\rule[-0.175pt]{4.818pt}{0.350pt}}
\put(421,756){\rule[-0.175pt]{4.818pt}{0.350pt}}
\put(599,621){\rule[-0.175pt]{0.350pt}{49.866pt}}
\put(589,621){\rule[-0.175pt]{4.818pt}{0.350pt}}
\put(589,828){\rule[-0.175pt]{4.818pt}{0.350pt}}
\put(766,684){\rule[-0.175pt]{0.350pt}{30.353pt}}
\put(756,684){\rule[-0.175pt]{4.818pt}{0.350pt}}
\put(756,810){\rule[-0.175pt]{4.818pt}{0.350pt}}
\put(1101,630){\rule[-0.175pt]{0.350pt}{41.194pt}}
\put(1091,630){\rule[-0.175pt]{4.818pt}{0.350pt}}
\put(1091,801){\rule[-0.175pt]{4.818pt}{0.350pt}}
\end{picture}
\caption{$m_0$ as a function of $n_f$ for $R_2 = 0.6$.} 
\end{center}
\end{figure}

\begin{figure} 
\begin{center}
\setlength{\unitlength}{0.240900pt}
\ifx\plotpoint\undefined\newsavebox{\plotpoint}\fi
\begin{picture}(1500,1619)(0,0)
\tenrm
\sbox{\plotpoint}{\rule[-0.175pt]{0.350pt}{0.350pt}}%
\put(264,832){\rule[-0.175pt]{282.335pt}{0.350pt}}
\put(264,158){\rule[-0.175pt]{4.818pt}{0.350pt}}
\put(242,158){\makebox(0,0)[r]{-0.25}}
\put(1416,158){\rule[-0.175pt]{4.818pt}{0.350pt}}
\put(264,293){\rule[-0.175pt]{4.818pt}{0.350pt}}
\put(242,293){\makebox(0,0)[r]{-0.2}}
\put(1416,293){\rule[-0.175pt]{4.818pt}{0.350pt}}
\put(264,428){\rule[-0.175pt]{4.818pt}{0.350pt}}
\put(242,428){\makebox(0,0)[r]{-0.15}}
\put(1416,428){\rule[-0.175pt]{4.818pt}{0.350pt}}
\put(264,562){\rule[-0.175pt]{4.818pt}{0.350pt}}
\put(242,562){\makebox(0,0)[r]{-0.1}}
\put(1416,562){\rule[-0.175pt]{4.818pt}{0.350pt}}
\put(264,697){\rule[-0.175pt]{4.818pt}{0.350pt}}
\put(242,697){\makebox(0,0)[r]{-0.05}}
\put(1416,697){\rule[-0.175pt]{4.818pt}{0.350pt}}
\put(264,832){\rule[-0.175pt]{4.818pt}{0.350pt}}
\put(242,832){\makebox(0,0)[r]{0}}
\put(1416,832){\rule[-0.175pt]{4.818pt}{0.350pt}}
\put(264,967){\rule[-0.175pt]{4.818pt}{0.350pt}}
\put(242,967){\makebox(0,0)[r]{0.05}}
\put(1416,967){\rule[-0.175pt]{4.818pt}{0.350pt}}
\put(264,1102){\rule[-0.175pt]{4.818pt}{0.350pt}}
\put(242,1102){\makebox(0,0)[r]{0.1}}
\put(1416,1102){\rule[-0.175pt]{4.818pt}{0.350pt}}
\put(264,1236){\rule[-0.175pt]{4.818pt}{0.350pt}}
\put(242,1236){\makebox(0,0)[r]{0.15}}
\put(1416,1236){\rule[-0.175pt]{4.818pt}{0.350pt}}
\put(264,1371){\rule[-0.175pt]{4.818pt}{0.350pt}}
\put(242,1371){\makebox(0,0)[r]{0.2}}
\put(1416,1371){\rule[-0.175pt]{4.818pt}{0.350pt}}
\put(264,1506){\rule[-0.175pt]{4.818pt}{0.350pt}}
\put(242,1506){\makebox(0,0)[r]{0.25}}
\put(1416,1506){\rule[-0.175pt]{4.818pt}{0.350pt}}
\put(264,158){\rule[-0.175pt]{0.350pt}{4.818pt}}
\put(264,113){\makebox(0,0){0}}
\put(264,1486){\rule[-0.175pt]{0.350pt}{4.818pt}}
\put(411,158){\rule[-0.175pt]{0.350pt}{4.818pt}}
\put(411,113){\makebox(0,0){1}}
\put(411,1486){\rule[-0.175pt]{0.350pt}{4.818pt}}
\put(557,158){\rule[-0.175pt]{0.350pt}{4.818pt}}
\put(557,113){\makebox(0,0){2}}
\put(557,1486){\rule[-0.175pt]{0.350pt}{4.818pt}}
\put(704,158){\rule[-0.175pt]{0.350pt}{4.818pt}}
\put(704,113){\makebox(0,0){3}}
\put(704,1486){\rule[-0.175pt]{0.350pt}{4.818pt}}
\put(850,158){\rule[-0.175pt]{0.350pt}{4.818pt}}
\put(850,113){\makebox(0,0){4}}
\put(850,1486){\rule[-0.175pt]{0.350pt}{4.818pt}}
\put(997,158){\rule[-0.175pt]{0.350pt}{4.818pt}}
\put(997,113){\makebox(0,0){5}}
\put(997,1486){\rule[-0.175pt]{0.350pt}{4.818pt}}
\put(1143,158){\rule[-0.175pt]{0.350pt}{4.818pt}}
\put(1143,113){\makebox(0,0){6}}
\put(1143,1486){\rule[-0.175pt]{0.350pt}{4.818pt}}
\put(1290,158){\rule[-0.175pt]{0.350pt}{4.818pt}}
\put(1290,113){\makebox(0,0){7}}
\put(1290,1486){\rule[-0.175pt]{0.350pt}{4.818pt}}
\put(1436,158){\rule[-0.175pt]{0.350pt}{4.818pt}}
\put(1436,113){\makebox(0,0){8}}
\put(1436,1486){\rule[-0.175pt]{0.350pt}{4.818pt}}
\put(264,158){\rule[-0.175pt]{282.335pt}{0.350pt}}
\put(1436,158){\rule[-0.175pt]{0.350pt}{324.733pt}}
\put(264,1506){\rule[-0.175pt]{282.335pt}{0.350pt}}
\put(-87,832){\makebox(0,0)[l]{\shortstack{{\Large $R_V(t)$}}}}
\put(850,23){\makebox(0,0){{\Large $t$}}}
\put(264,158){\rule[-0.175pt]{0.350pt}{324.733pt}}
\put(704,834){\makebox(0,0){$\star$}}
\put(850,885){\makebox(0,0){$\star$}}
\put(997,872){\makebox(0,0){$\star$}}
\put(1143,898){\makebox(0,0){$\star$}}
\put(1290,955){\makebox(0,0){$\star$}}
\put(704,791){\rule[-0.175pt]{0.350pt}{20.717pt}}
\put(694,791){\rule[-0.175pt]{4.818pt}{0.350pt}}
\put(694,877){\rule[-0.175pt]{4.818pt}{0.350pt}}
\put(850,788){\rule[-0.175pt]{0.350pt}{46.735pt}}
\put(840,788){\rule[-0.175pt]{4.818pt}{0.350pt}}
\put(840,982){\rule[-0.175pt]{4.818pt}{0.350pt}}
\put(997,672){\rule[-0.175pt]{0.350pt}{96.360pt}}
\put(987,672){\rule[-0.175pt]{4.818pt}{0.350pt}}
\put(987,1072){\rule[-0.175pt]{4.818pt}{0.350pt}}
\put(1143,529){\rule[-0.175pt]{0.350pt}{177.784pt}}
\put(1133,529){\rule[-0.175pt]{4.818pt}{0.350pt}}
\put(1133,1267){\rule[-0.175pt]{4.818pt}{0.350pt}}
\put(1290,207){\rule[-0.175pt]{0.350pt}{312.929pt}}
\put(1280,207){\rule[-0.175pt]{4.818pt}{0.350pt}}
\put(1280,1506){\rule[-0.175pt]{4.818pt}{0.350pt}}
\end{picture}
\caption{The ratio $R_V(t)$ for $n_f = -4$; $R_2 = 0.5$.} 
\end{center}
\end{figure}
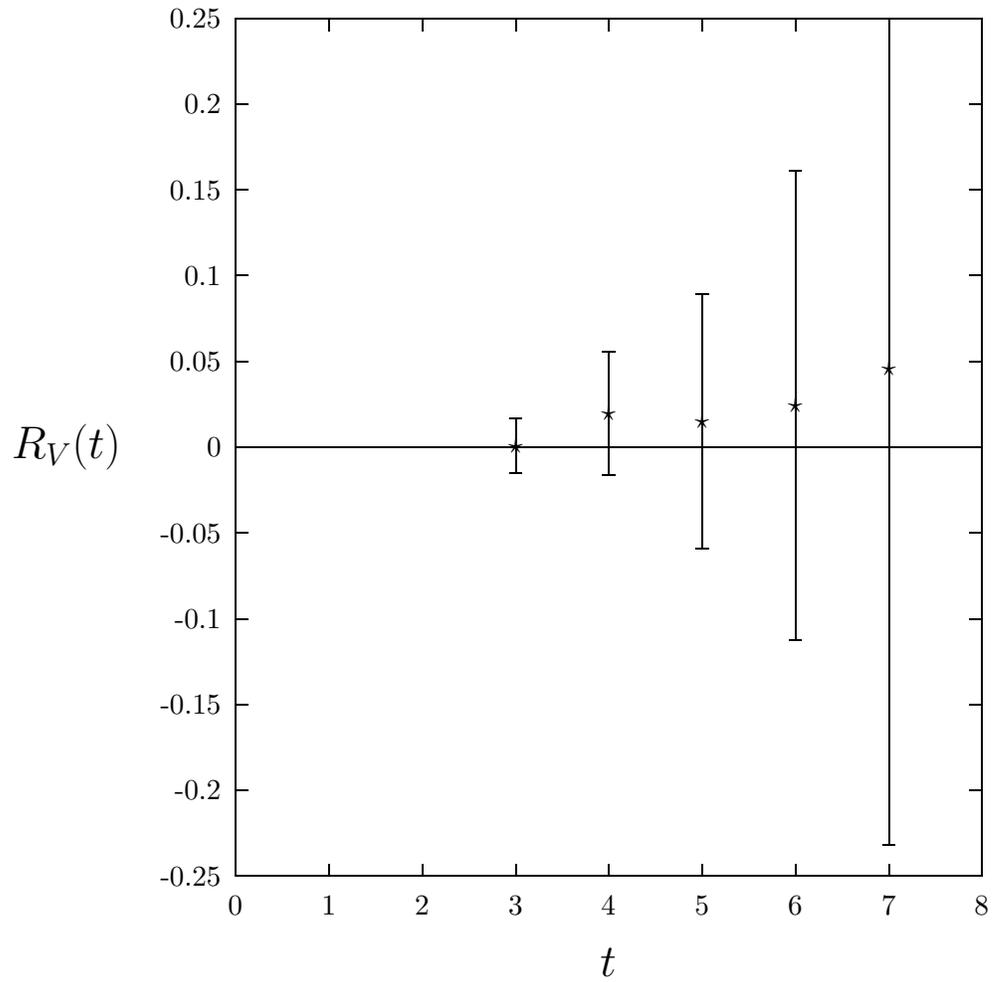

\end{document}